 \documentclass[aps,showpacs,showkeys,amsmath,amssymb,preprint,prd,%
 floatfix,superscriptaddress]{revtex4}
 
 \usepackage{graphicx}
 \usepackage{dcolumn}
 \usepackage{docs}%
 \usepackage{bm}
 \begin{document}
 \title{Quasiparticle Model of Quark-Gluon Plasma \\ 
 at Imaginary Chemical Potential}
 \author{M.~Bluhm}
 \affiliation{
 Forschungszentrum Dresden-Rossendorf, PF 510119, 01314 Dresden, Germany}
 \author{B.~K\"ampfer}
 \affiliation{
 Forschungszentrum Dresden-Rossendorf, PF 510119, 01314 Dresden, Germany}
 \affiliation{Institut f\"ur Theoretische Physik, TU Dresden, 
 01062 Dresden, Germany}
 
 \date{\today}
 
 
 \begin{abstract}
 A quasiparticle model of the quark-gluon plasma is
 compared with lattice QCD data for purely imaginary chemical
 potential. Net quark number density, 
 susceptibility as well as the deconfinement border 
 line in the phase diagram of strongly 
 interacting matter are investigated. In addition, the 
 impact of baryo-chemical potential dependent 
 quasiparticle masses is discussed. This accomplishes 
 a direct test of the model for non-zero baryon density. 
 The found results are 
 compared with lattice QCD data for real 
 chemical potential by means of analytic continuation 
 and with a different (independent) set 
 of lattice QCD data at zero chemical potential. 
 \end{abstract}
 
 \maketitle
 
 \section{Introduction}
 
 Strongly interacting matter, as described by QCD, exhibits an
 astonishingly rich phase structure. In the region of not too large
 baryon densities, the deconfinement transition from hadronic matter
 to a plasma built of quark-gluon constituents is the most prominent
 feature. It is signalled by a rapid change in the expectation value
 of the Polyakov loop and the chiral condensate where one assigns a
 pseudo-critical temperature $T_c$ to this transition 
 (cf. reviews, e.~g.~\cite{Karsch}). At higher
 temperatures, $T > 3 T_c$, further structural changes are conjectured
 \cite{Hatsuda}. For non-zero quark chemical potential $\mu$, corresponding
 to a finite net baryon density, many researchers argue on the change of
 the deconfinement border line, representing an analytic 
 crossover, into a first-order transition curve. The onset of this 
 sequence of first-order transitions is marked by a critical point 
 being of second order which has attracted 
 much attention recently (see~\cite{POSCPOD}). The interest in
 this part of the phase diagram is triggered by the possibility to
 probe it under laboratory conditions in relativistic heavy-ion
 collisions.
 
 With the advance of precision data from ultrarelativistic 
 heavy-ion collisions at RHIC, the paradigm on the
 quark-gluon plasma has changed~\cite{Gyulassy}: The notion of a strongly
 coupled plasma has been put forward to explain the seemingly very
 small viscosity to entropy ratio deduced from hydrodynamical fits to 
 experimental data as in~\cite{expdata}, and various models have
 been developed \cite{SQGP} to account for such a property. On the other hand, we
 are witnessing a vast progress in first principle calculations of
 thermodynamic properties of hot deconfined strongly interacting matter 
 based directly on QCD \cite{Laermann,Philipsen,C_Schmidt}. While
 various observables such as pressure, energy density or numerous
 susceptibilities are addressed, the available lattice QCD data
 are obtained for different numerical set-ups, lattice sizes, flavor
 numbers and quark masses as well. Particular attempts are needed to
 access non-zero baryon densities because the notorious sign problem of the
 fermion determinant prevents a direct application of methods useful
 for zero baryon density. Nevertheless, a few methods have been
 developed to access non-zero baryon densities. Among such
 methods is the calculation of thermodynamic quantities at
 purely imaginary chemical potential. Here, the sign problem is
 avoided but the results have to be analytically continued to real
 chemical potential. In this respect it is useful to have a model at
 our disposal which is successfully probed for both, real and
 imaginary chemical potential, in order to accomplish the translation of
 results from imaginary to real chemical potential.
 
 While baryon density effects are small for heavy-ion collisions at
 top-RHIC energies and will be even smaller for LHC energies, at
 least in the mid-rapidity region, for CERN-SPS and upcoming FAIR
 energies they are sizeable. In this respect, a firm knowledge of
 thermodynamic bulk properties of strongly interacting matter is
 highly desirable. As a step towards achieving this goal we are going to extend
 our quasiparticle model~\cite{Peshier1,Peshier3,MBDipl,Bluhm-EPJC} to imaginary
 chemical potential. Here, information is obtained for 
 $\mu^2 < 0$ allowing, in principle, for identifying 
 $\mu=\pm i\mu_i$. The model has been tested successfully for real
 chemical potential \cite{Bluhm-PLB}, say in describing the Taylor
 expansion coefficients of the pressure as a series in powers of
 $\mu / T$. In such a way,
 Peshier's flow equation \cite{Peshier3} is tested in some detail.
 This flow equation transports information about the effective
 coupling, $G^2$, from the temperature axis to non-zero $\mu$ and
 determines to a large extent the dependence on $\mu$ and thus on
 baryon density. Another important piece of the model is the
 quasiparticle ansatz for dynamically generated effective masses of
 quarks $\propto (T^2+\frac{\mu^2}{\pi^2})G^2$. When going to purely
 imaginary chemical potential $\mu \rightarrow \pm i\mu$ the sign
 of the $\mu^2$ term is flipped, as also signs in Peshier's flow equation
 are changed. Therefore, the $\mu$ dependence of the model is directly 
 tested by considering an imaginary chemical potential.
 
 In the following, two symmetries of the QCD partition function
 $Z(T,\mu)$ are of relevance: 
 (i) $Z(T,\mu) = Z(T,-\mu)$, and 
 (ii) $Z(T,i\mu_i) = Z(T,i(\mu_i+\frac{2\pi}{3}T))$, i.~e.~$Z(T,\mu)$ 
 is periodic in $\mu_i$ with period $2\pi T/3$~\cite{RW}. 
 Symmetry (i) makes $Z$ an
 even function of $\mu$ (meaning that in a Taylor series expansion 
 only even powers of $\mu/T$, and thus also of $\mu_i/T$, are encountered) 
 such that we can focus on $\mu = +i\mu_i$ only, while (ii) implies the
 Roberge-Weiss periodicity~\cite{RW} which is anchored in the 
 center symmetry. This periodicity is characterized by 
 lines of first-order transitions ($Z_3$
 transitions) at $\mu_i = \frac{\pi}{3} T(1+2k)$ for all integers $k$ and 
 sufficiently high temperature $T$ while for smaller temperatures 
 the behavior of thermodynamic quantities is analytic. The endpoint of 
 first-order transitions, $T^E$, is determined by the crossing of the 
 Roberge-Weiss transition line with the chiral critical line which is 
 also a first-order transition line for $N_f=4$ degenerate 
 quark flavors~\cite{MPL2}. 
 The Roberge-Weiss periodicity implies that in the $T-\mu_i/T$ plane all sectors between
 $\mu_i/T =\frac{2\pi}{3}k$ and $\mu_i/T =\frac{2\pi}{3}(k+1)$ are copies of the
 sector between $\mu_i/T =0$ and $\mu_i/T =\frac{2\pi}{3}$. Furthermore, the subsector between 
 $\mu_i/T =\pi/3$ and $\mu_i/T =2\pi/3$ is an reflected copy of the subsector between 
 $\mu_i/T =0$ and 
 $\mu_i/T =\pi/3$ mirrored at the first Roberge-Weiss transition line. 
 As thermodynamic quantities behave non-analytically at $\mu_i/T = \pi/3$ 
 (Roberge-Weiss transition), an
 analytic continuation of results obtained for imaginary 
 chemical potential to real $\mu$ has direct access to the region 
 $\mu < \pi/3\,T $ only. 
 
 Due to the severe approximations made when linking our 
 phenomenological model~\cite{Peshier1,Peshier3,MBDipl} 
 to QCD \cite{Bluhm-EPJC}, the Roberge-Weiss periodicity is not longer 
 apparent. 
 Having this in mind, we translate the model to imaginary
 chemical potential in section~\ref{sec:2}. The comparison with lattice QCD data at
 imaginary chemical potential is performed in section~\ref{sec:3}, where also
 the continuation to real chemical potential is presented. This allows, in 
 addition, for a comparison with another and independent set of 
 lattice QCD data obtained at $\mu=0$ (section~\ref{sec:4}). 
 Furthermore, we investigate in detail the impact of the baryo-chemical 
 potential dependence of the quasiquark and quasigluon masses 
 (selfenergies) on the found results and discuss the deconfinement border 
 line in the phase diagram of strongly 
 interacting matter. Our results are summarized in section~\ref{sec:5}. 
 Appendices A and B contain Peshier's flow equation for 
 imaginary chemical potential and a discussion about the 
 parametrization of the $\mu$ dependence of the density. 
 
 \section{Quasiparticle model at imaginary chemical potential \label{sec:2}}
 
 The employed model is based on a two-loop {\boldmath$\Phi$} functional 
 approach to QCD with corresponding one-loop self-energies 
 considered in HTL approximation in the asymptotic limit 
 and the neglect of finite width effects, (anti)plasmino 
 and longitudinal gluon contributions as well as Landau 
 damping \cite{Bluhm-EPJC}. The QCD running 
 coupling is replaced by an effective coupling $G^2(T,\mu)$ 
 which is subject to Peshier's flow equation~\cite{Peshier3} resting on a thermodynamic
 self-consistency condition and the stationarity of the grand canonical
 potential $\Omega=-pV=-T\ln Z$, 
 where $p$ denotes the pressure and $V$ the volume of the system. 
 Straightforward replacement of $\mu=i\mu_i$ in $p(T,\mu)$ renders the 
 net quark number density, 
 $n(T,i\mu_i)=-i\partial p(T,i\mu_i)/\partial\mu_i$, related to
 the net baryon density $n_B = \frac13 n$, to 
 \begin{eqnarray}
   \label{equ:dens1}
   n(T,i\mu_i) & = & \frac{d_q}{2\pi^2} \int_0^\infty dk k^2  
   \left( \frac{1}{e^{(\omega_q-i\mu_i)/T}+1} -
   \frac{1}{e^{(\omega_q+i\mu_i)/T}+1} \right) \\
   \label{equ:dens2}
   & = & i\frac{d_q}{\pi^2}\int_0^\infty dk k^2 \left(\frac{e^{\omega_q/T}\sin (\mu_i/T)}
   {e^{2\omega_q/T}+2e^{\omega_q/T}\cos (\mu_i/T)+1}\right) , 
 \end{eqnarray}
 where $d_q=2N_cN_f$ is the degeneracy factor of quarks for $N_c=3$ 
 colors and $N_f$ quark flavors. The found result for $n$ is purely 
 imaginary and positive (negative) for small positive (negative) $\mu_i$, 
 i.~e.~$n$ is an odd function in $\mu_i$. 
 
 The quark dispersion relation $\omega_q(k)$ employed in Eqs. 
 (\ref{equ:dens1},~\ref{equ:dens2}) reads 
 \begin{equation}
   \label{equ:dispq}
   \omega_q^2 = k^2 + M_\infty^2 
 \end{equation}
 with asymptotic mass $M_\infty^2 = m_q^2 +2M_+^2$ using 
 \begin{equation}
 \label{equ:plafreq}
 M_+^2 = \frac{N_c^2-1}{16N_c}(T^2 - \frac{\mu_i^2}{\pi^2}) G^2 (T,i\mu_i)
 \end{equation}
 as plasma frequency. This dispersion relation is based on a 
 calculation of one-loop self-energies with finite quark masses 
 $m_q$ in Feynman gauge in the asymptotic limit 
 \cite{Seipt_Diploma} for small $m_q/T$, where $m_q$ 
 may be temperature dependent as well to allow for a direct 
 comparison with lattice QCD data. 
 (A different approximation of $M_\infty$ is discussed in 
 section~\ref{sec:3_3}.) Eqs.~(\ref{equ:dens1},~\ref{equ:dens2}) highlight the 
 quasiparticle character of the model: the
 baryon charge is carried by excitations with dispersion relation given by
 Eq.~(\ref{equ:dispq}). The dependence of $M_\infty$ 
 on the chemical potential (cf.~\cite{Shuryak}) 
 will be discussed in section~\ref{sec:3_4}. 
 
 Peshier's flow equation~\cite{Peshier3} for imaginary chemical potential reads 
 \begin{equation}
   \label{equ:PDE}
   b = a_T \frac{\partial G^2}{\partial T} + a_{\mu_i} 
   \frac{\partial G^2}{\partial\mu_i} \,, 
 \end{equation}
 where the coefficients $b$, $a_T$ and $a_{\mu_i}$ depending on 
 $T$, $\mu_i$ and $G^2(T,i\mu_i)$ are relegated to 
 Appendix A. Transforming Eq. (\ref{equ:PDE}) to a system of 
 three coupled ordinary differential equations, it can be solved 
 by the methods of characteristics knowing, for instance, 
 $G^2(T,\mu=0)$. A convenient parameterization of $G^2(T,\mu=0)$ 
 is~\cite{Peshier3} 
 \begin{equation}
 \label{equ:coupl}
 G^2(T\ge T_c,\mu=0) = \frac{16 \pi^2}{\beta_0 \log \xi^2} \,,
 \end{equation}
 making some contact to perturbative QCD at very large 
 temperatures. 
 Here, $\beta_0 = \frac{1}{3}(11 N_c - 2 N_f)$
 and $\xi$ is parametrized 
 phenomenologically as $\xi = \lambda (T - T_s)/T_c$ with scale 
 parameter $\lambda$ and $T_s$ shifting the infrared divergence 
 to $T=T_s+T_c/\lambda < T_c$ for appropriate parameters, 
 while we focus here on the region $T\ge T_c$. 
 
 Results obtained for $\mu^2<0$ need to be analytically continued 
 into the $\mu^2>0$ half-plane in order to achieve physical results. 
 An effective analytic continuation requires a positive second 
 derivative of $Z$ with respect to $\mu$, 
 cf.~\cite{Lombardo1,MPL3}, i.~e.~the quark 
 number susceptibility 
 $\chi(T,\mu) = \partial n(T,\mu)/\partial\mu > 0$. 
 The QPM result for $\chi$ reads for imaginary chemical potential 
 \begin{eqnarray}
   \label{e:susc}
   \nonumber
   \chi(T,i\mu_i) & = & \frac{d_q}{2\pi^2 T} \int_0^\infty dk k^2 
   \frac{\left( 2e^{3\omega_q/T}\cos (\mu_i/T) 
      + 4e^{2\omega_q/T} + 2e^{\omega_q/T}\cos (\mu_i/T)\right)}
      {\left(e^{2\omega_q/T} + 2e^{\omega_q/T} 
      \cos (\mu_i/T) + 1\right)^2} \\ 
   & & + \frac{d_q}{2\pi^2 T} \int_0^\infty dk 
   \frac{k^2}{\omega_q}\frac{\left(e^{3\omega_q/T} \sin (\mu_i/T) - 
      e^{\omega_q/T}\sin (\mu_i/T)\right)}{\left(e^{2\omega_q/T} + 2e^{\omega_q/T} 
      \cos (\mu_i/T) + 1\right)^2} \\ \nonumber
   & & \times \frac{N_c^2-1}{8N_c}\left(\frac{2}{\pi^2}\mu_i G^2 - \left[T^2 - 
      \frac{\mu_i^2}{\pi^2}\right]\frac{\partial G^2}{\partial \mu_i}\right) \,;
 \end{eqnarray}
 it is purely real and symmetric under 
 $\mu_i\rightarrow -\mu_i$ (cf.~Appendix A). Furthermore, 
 for small $\mu_i$, the first term in Eq.~(\ref{e:susc}) is positive 
 and dominates the second term. At $\mu=0$, one finds  
 \begin{equation}
   \label{e:susc1}
   \chi(T,\mu=0) = \frac{d_q}{\pi^2 T}\int_0^\infty dk k^2
    \frac{e^{\tilde{\omega}_q/T}}{e^{2\tilde{\omega}_q/T}+
    2e^{\tilde{\omega}_q/T}+1} > 0 \,,
 \end{equation}
 where $\tilde{\omega}_q=\omega_q(T,\mu=0)$. 
 
 \section{Comparison with lattice QCD results for imaginary 
 mu \label{sec:3}}
 
 \subsection{Baryon density and quark number susceptibility \label{sec:3_1}}

 We confront now the above introduced quasiparticle 
 model (QPM) with lattice QCD data~\cite{Lombardo1,MPL} 
 at non-zero $T$ and $\mu_i$ 
 obtained for $N_f=4$ degenerate quark flavors with 
 $m_q=0.2\,T$; these calculations~\cite{Lombardo1,MPL} 
 are performed on a lattice with temporal and spatial 
 extensions $N_\tau=4$ and $N_\sigma=16$. Valuable information 
 in such simulations with imaginary chemical potential is obtained for 
 $\mu_i < 0$~\cite{MPL-X} implying a negative 
 imaginary part of the net quark number 
 density $n(T,i\mu_i)$ according to Eq.~(\ref{equ:dens2}). 
 In the following, however, we will 
 consider $\mu_i > 0$ which renders the sign of $n(T,i\mu_i)$ and 
 accordingly the behavior of $\chi=\partial n/\partial (i\mu_i)$. 
 Our model is formulated for a system infinite in space and time. 
 Thus, we need a proper extrapolation of the lattice QCD data to the 
 continuum limit ($N_\tau\rightarrow\infty$ at fixed temperature). 
 Different estimates for a continuum 
 extrapolation are conceivable. For instance, one may 
 select a scaling factor strictly valid only for 
 asymptotically high temperatures, 
 or one may use as scaling factor the ratio of thermodynamic quantities for a massless, 
 non-interacting gas of quarks and gluons known in the continuum 
 limit and from lattice QCD for finite $N_\tau$. 
 Even though such estimates for a correction factor 
 could depend on $T$, in 
 general, we apply the latter procedure, assuming that the 
 continuum extrapolations for QCD and for the non-interacting 
 gas of quarks and gluons are similar (cf.~discussion in \cite{latSU3} 
 for pure SU(3) theory). In principle, however, a profound extrapolation 
 to the continuum limit should be based on simulations with different 
 size lattices as leading corrections to the continuum limit are of the 
 order $\mathcal{O}(N_\tau^{-2})$~\cite{latSU3,Karsch_overview}. 
 Taking the Stefan-Boltzmann result 
 of $n_B/T^3$ for $N_\tau=4$ \cite{Lombardo1}, we find as educated 
 guess for the needed continuum extrapolation factor of the 
 net quark number density $d_{lat}^{(n)}=0.456$. This compares well 
 with continuum extrapolation factors reported in 
 \cite{Szabo,Gavai}, reading $0.446$ and $0.465$ respectively, where 
 similar actions have been used in the lattice simulations. 
 
 In Fig.~\ref{fig.1}, we compare our model with the continuum estimate of 
 the lattice QCD data \cite{MPL,Lombardo1} for the scaled 
 net quark number density as a function of $\mu_i/T_c$ at 
 constant $T$. 
 \begin{figure}[t]
   \includegraphics[scale=0.35,angle=0.]{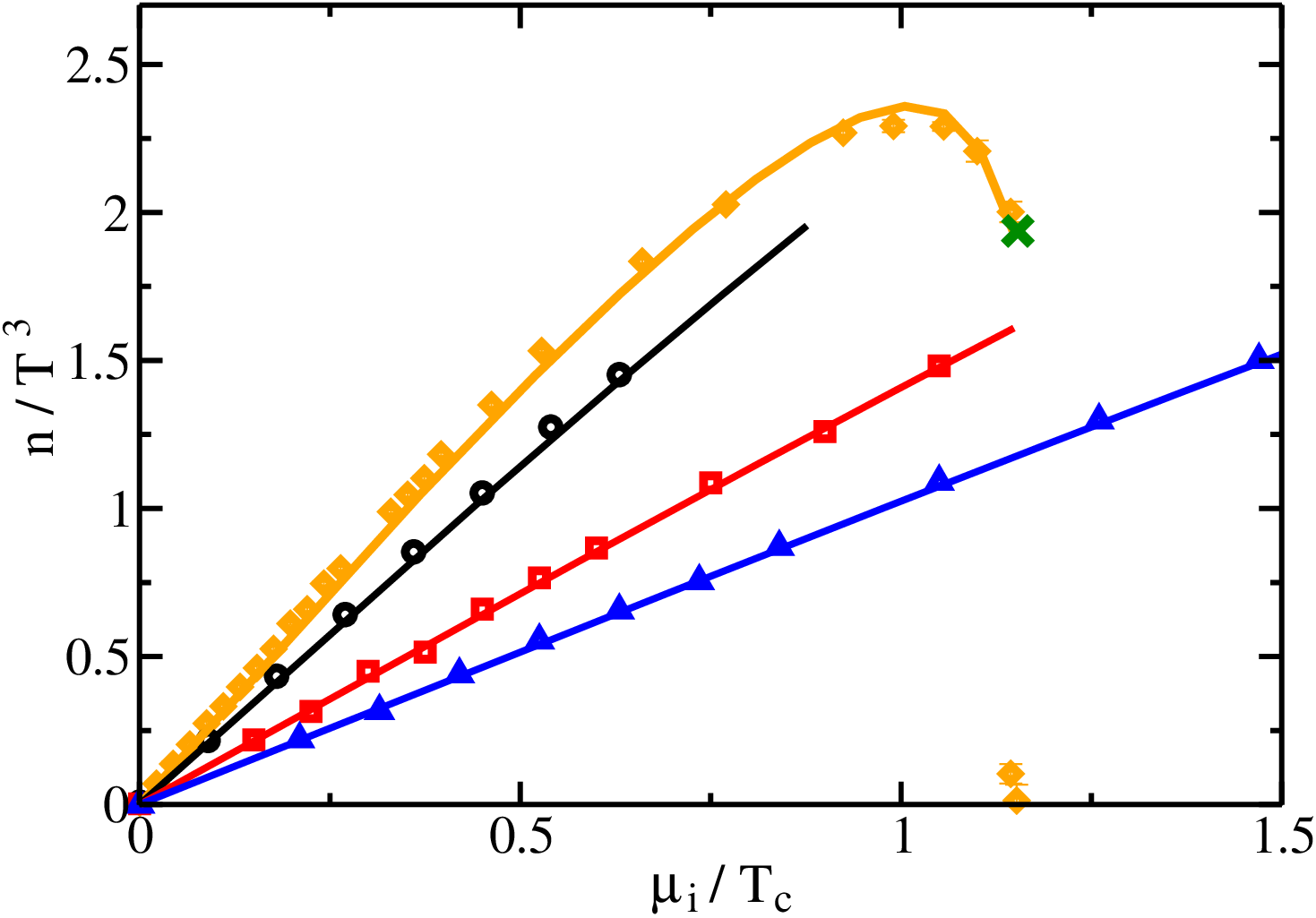}
   \caption{Comparison of the QPM (solid curves) for 
    the scaled net quark number density $n/T^3$ as a 
    function of $\mu_i/T_c$ 
    with continuum estimates of the lattice QCD data~\cite{Lombardo1,MPL} 
    for temperatures $T=1.1, 1.5, 2.5, 3.5\,T_c$ (diamonds, 
    circles, squares and triangles, respectively). 
    The fat cross 
    depicts the Roberge-Weiss critical chemical potential 
    $\mu_c/T=\pi/3$ for $T=1.1\,T_c$, where we stopped our 
    calculations. 
    \label{fig.1}}
 \end{figure}
 Because $n=3\,n_B$ as a function of imaginary chemical potential 
 is found to be purely imaginary, both in Eq.~(\ref{equ:dens2}) and 
 in the lattice calculations, we exhibit its imaginary part 
 in the following. 
 The parameters of the effective coupling 
 $G^2(T,\mu=0)$ in Eq.~(\ref{equ:coupl}) read $T_s=0.96\,T_c$ 
 and $\lambda=56$ shifting the divergence 
 of $G^2(T,\mu=0)$ to approximately $T=0.98\,T_c$, 
 where we utilize $T_c = 163$~MeV 
 as given in \cite{MPL2} for the case at hand. Note that we consider 
 only temperatures $T\ge T_c$. The continuum extrapolated lattice QCD 
 data, in particular the pronounced bending of $n/T^3$ for 
 $T=1.1\,T_c$, are impressively well described by the QPM 
 parametrization. The drastic change in the slope for 
 $T=1.1\,T_c$ signals the 
 onset of the Roberge-Weiss transition at 
 $\mu_c/T_c=11\pi/30$, where $n$ should exhibit 
 a discontinuity. In the QPM, this change in slope is 
 driven by the dependence of the quasiparticle asymptotic mass 
 $M_\infty$ on chemical potential and, in particular, 
 by the behavior of $G^2$ with respect to $\mu_i$ as dictated by 
 Peshier's flow equation Eq.~(\ref{equ:PDE}). We note that 
 $n/T^3$ exhibited as a function of $\mu_i/T$ shows almost no 
 dependence on $T$ for temperatures $T\ge 1.5\,T_c$. Below $T_c$, 
 however, $n/T^3$ displays a qualitatively different behavior 
 being continuous and periodic as a function of 
 $\mu_i/T$~\cite{Lombardo1}. 
 
 Within the QPM, results obtained by considering purely imaginary 
 chemical potential can easily be analytically continued to 
 real $\mu$. 
 \begin{figure}[t]
   \includegraphics[scale=0.35,angle=0.]{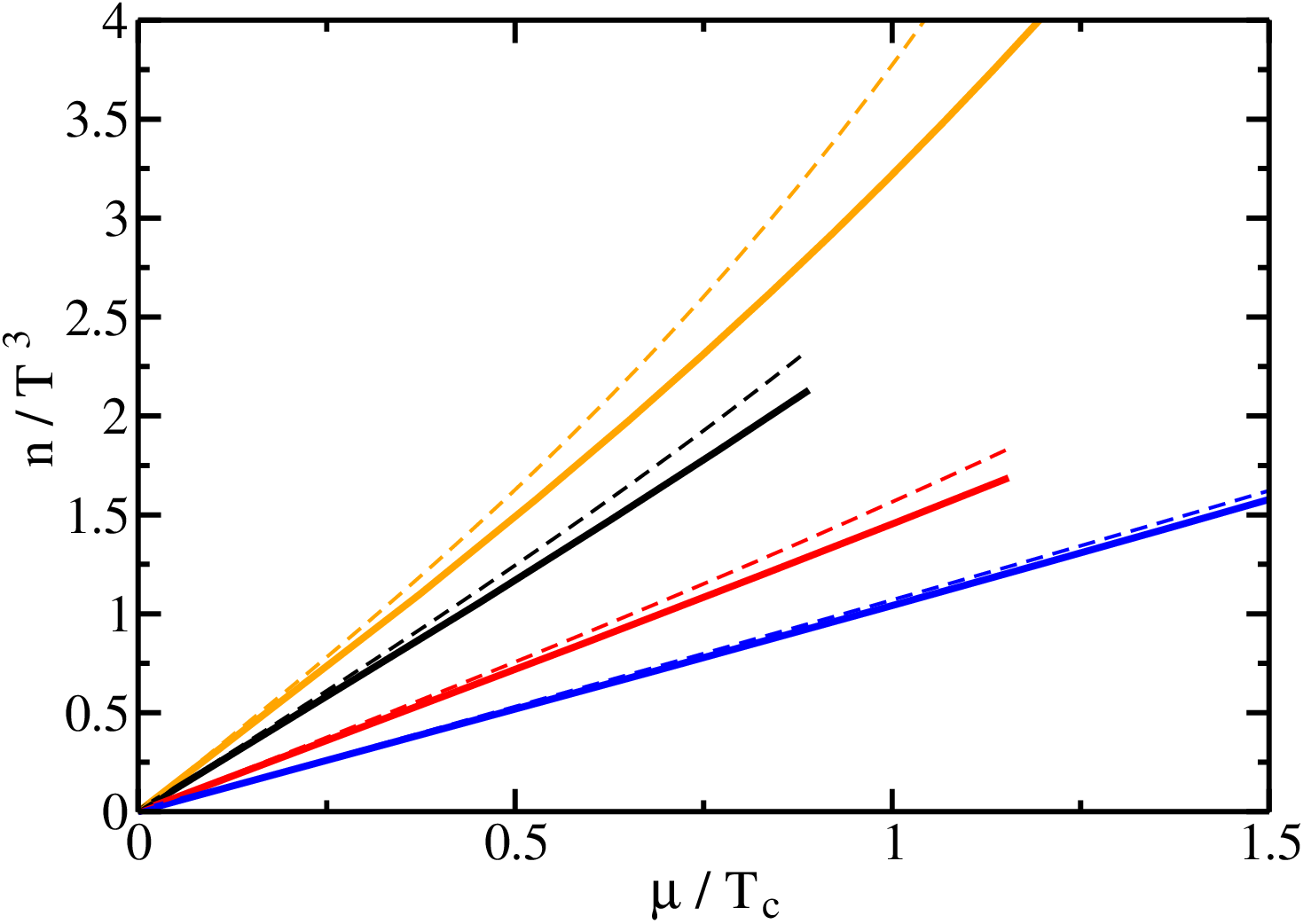}
   \caption{
     Continuation of the QPM results for $n/T^3$ 
     exhibited in Fig.~\ref{fig.1} to 
     real chemical potential $\mu/T_c$ (solid curves) 
     for $T=1.1, 1.5, 2.5, 3.5\,T_c$ (from top to bottom). 
     For comparison, we also show the 
     analytically continued results (dashed curves) 
     of the polynomial fit from~\cite{Lombardo1} to $n/T^3$ 
     for imaginary chemical potential. \label{fig.3}}
 \end{figure}
 This is achieved by continuing the purely imaginary variable 
 $\mu=i\mu_i$ to the entire complex plane and finally taking the limit 
 ${\rm Im}\,\mu\rightarrow 0$. In this way, we recover the quasiparticle 
 model~\cite{Peshier1,Peshier3,Bluhm-EPJC} formulated for 
 real $\mu$. Within the analyticity domain, i.~e.~for $\mu < \mu_c(T)$, 
 the analytic continuation is unique as guaranteed by general 
 arguments. 
 Keeping the QPM parameters $\lambda$ and $T_s$ fixed, the 
 results of $n/T^3$ for real $\mu/T_c$ are exhibited in Fig.~\ref{fig.3} 
 (solid curves). These results may be compared to other 
 analytic continuations. For instance, 
 in~\cite{Lombardo1}, a polynomial fit to $n/T^3$ as 
 well as its analytic continuation to real $\mu/T$ 
 (dashed curves in Fig.~\ref{fig.3}) 
 was considered. Despite the fact 
 that this polynomial fit $n(T,\mu_i,m_q)=a(T,m_q)\mu_i+b(T,m_q)\mu_i^3$ 
 for imaginary chemical potential, 
 with analytic continuation $n(T,\mu,m_q)=a(T,m_q)\mu-b(T,m_q)\mu^3$, 
 cannot account for the change in slope 
 observed for $T=1.1\,T_c$ at large $\mu_i/T_c$, its coefficients $a$ and $b$ 
 are temperature and quark mass dependent. In contrast, the 
 QPM parameters $\lambda$ and $T_s$ are once adjusted to $n/T^3$ 
 at $T=1.1\,T_c$ (cf.~Fig.~\ref{fig.1}) and then kept fixed for 
 all temperatures and chemical potentials. In addition, 
 the behavior of analytic continuations of polynomial fits 
 decisively depends on the considered order in $\mu_i^2$ (cf.~discussion 
 in~\cite{Lombardo3,C_Schmidt}). The QPM, in contrast, contains 
 all orders of $\mu_i^2$ respecting the symmetry 
 $\ln Z(\mu) = \ln Z(-\mu)$. As evident from Fig.~\ref{fig.3}, 
 we point out that close to $\mu_c(T)$ a sensible 
 analytic continuation is needed. 
 
\begin{figure}[t]
   \includegraphics[scale=0.35,angle=0.]{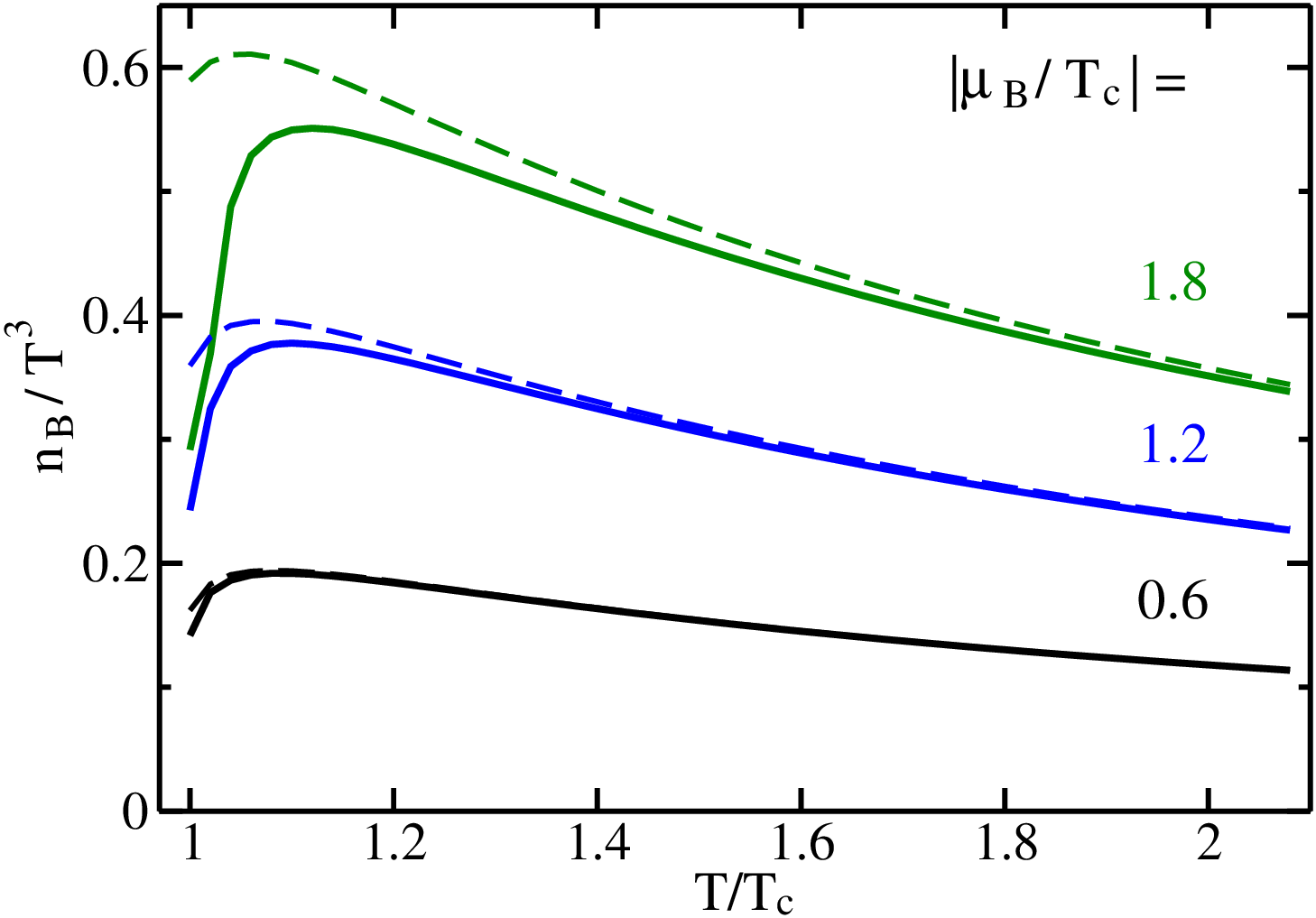}
   \caption{Scaled net baryon density $n_B/T^3$ 
     as a function of $T/T_c$ for constant 
     imaginary $\mu_B/T_c=3i\mu_i/T_c$ (solid curves) 
     and for corresponding real $\mu_B/T_c$ (dashed 
     curves). Note that for all temperatures, $\partial n/\partial T > 0$ 
     is fulfilled, 
     as required from thermodynamic stability conditions. \label{fig.4}}
 \end{figure}

 In Fig.~\ref{fig.4}, the net baryon density $n_B/T^3$ is 
 exhibited as a function of $T/T_c$ for constant 
 imaginary (solid curves) as well 
 as for real baryo-chemical potential $\mu_B=3\,\mu$ 
 (dashed curves). 
 Somewhat surprisingly, the results 
 for real $\mu_B$ significantly 
 deviate from the original results for imaginary chemical potential 
 only at large $\mu_B$ and temperatures close to $T_c$. 
 Note that in these considerations $\mu_B$ is 
 restricted to $|\mu_B|\le \pi T$. 
 
\begin{figure}[t]
   \includegraphics[scale=0.3,angle=0.]{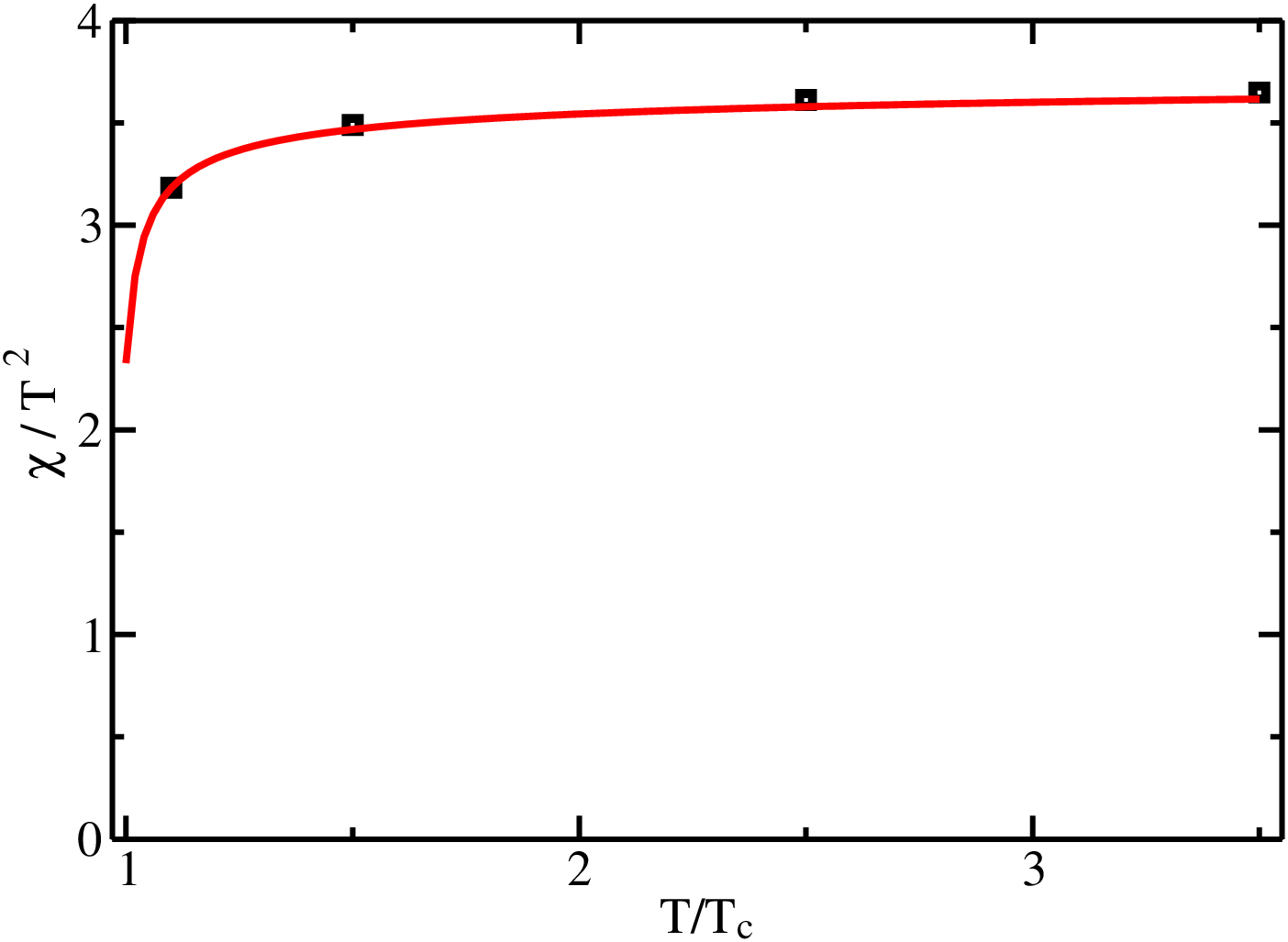}
   \includegraphics[scale=0.3,angle=0.]{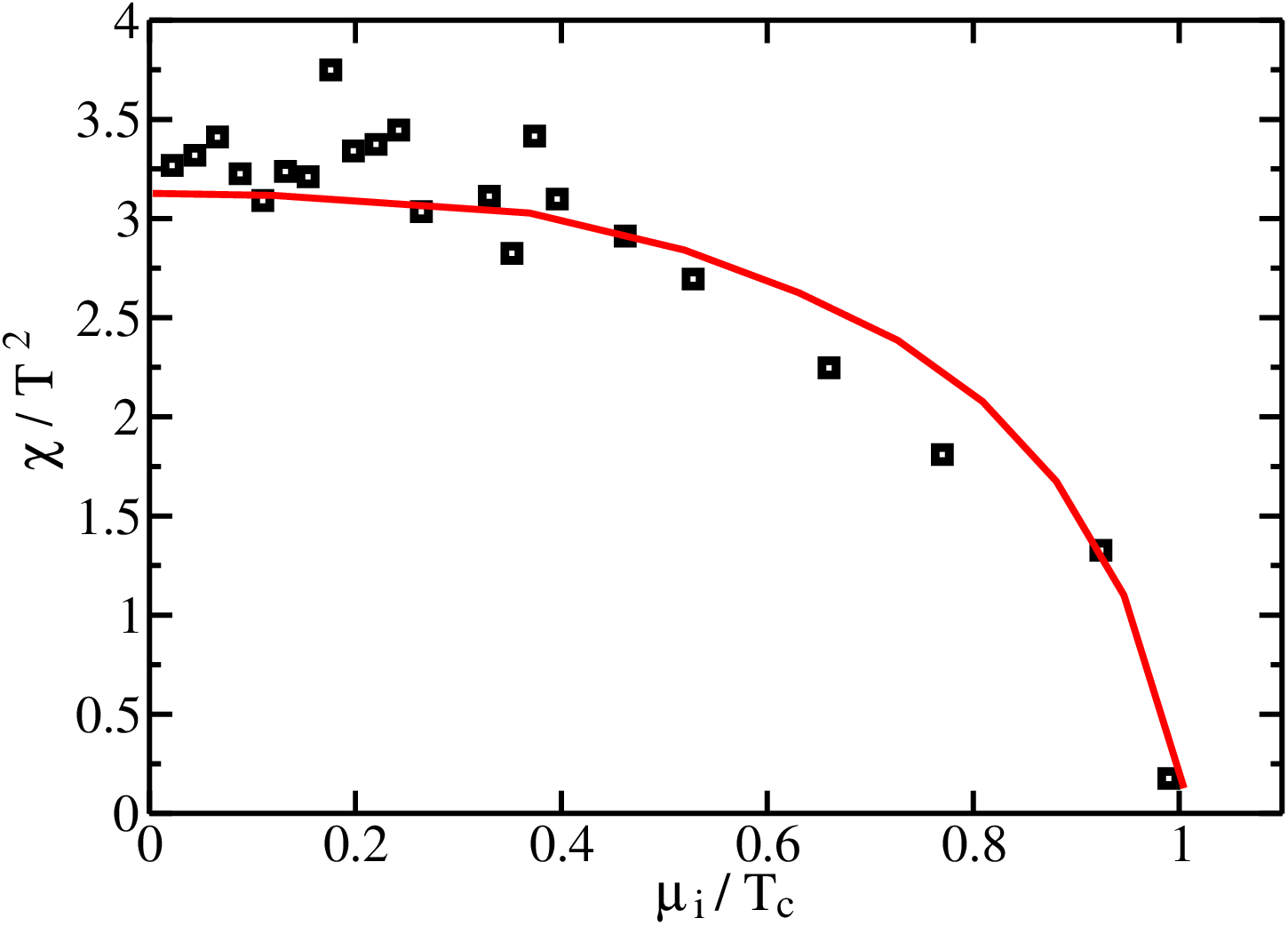}
   \caption{Left: comparison of the QPM (solid curve) for the 
   scaled quark number susceptibility $\chi/T^2$ 
   as a function of $T/T_c$ for $\mu=0$ with the continuum 
   estimate of the lattice QCD data in~\cite{Lombardo1} (circles). 
   Right: comparison of the QPM (solid curve) 
   for $\chi/T^2$ as a function 
   of $\mu_i/T_c\le 1$ for $T=1.1\,T_c$ with the continuum 
   estimate of the lattice QCD data in~\cite{MPL}. \label{fig.B9}}
 \end{figure}

 Susceptibilities are quantities serving as measures of 
 fluctuations. The quark number susceptibility 
 $\chi$ (cf.~Eqs.~(\ref{e:susc},~\ref{e:susc1})) 
 is simply the derivative of the density in $\mu_i$ direction. 
 We exhibit $\chi/T^2$ either at $\mu=0$ for various temperatures (Fig.~\ref{fig.B9}, 
 left panel) or for $T=1.1\,T_c$ for various values of $\mu_i$ 
 (Fig.~\ref{fig.B9}, right panel). 
 Clearly, if lattice QCD data for $n(T,i\mu_i)$ are well described by a model, 
 the model should also describe $\chi(T,i\mu_i)$. This is indeed the 
 case, see Fig.~\ref{fig.B9}, where both, lattice QCD data as well as 
 QPM results, are obtained by numerical differentiation of the net 
 quark number density. The only concern that could arise 
 is that derivatives enhance possible systematic differences 
 between a model and the data. Fig.~\ref{fig.B9} does not point 
 to such a possibility. 
 
 
  
 \subsection{Deconfinement border line \label{sec:3_2}}
 
 The solution of Peshier's flow equation 
 Eq.~(\ref{equ:PDE}) is accomplished by the method
 of characteristics. As in \cite{Peshier3}, we consider the
 characteristic curve emerging at $T = T_c$ and $\mu = 0$ 
 as an indicator of the pseudo-critical line. This 
 transition line has been calculated in lattice simulations 
 \cite{MPL2} for imaginary chemical potential. 
 The lattice QCD data have been analyzed by applying 
 polynomial fits which were analytically continued to real 
 $\mu$~\cite{MPL2}. The results of such analytic continuations 
 decisively depend on the chosen degree of the 
 considered polynomial or ratios thereof, as discussed in 
 \cite{C_Schmidt,Lombardo3}. 
 
\begin{figure}[t]
   \includegraphics[scale=0.3,angle=0.]{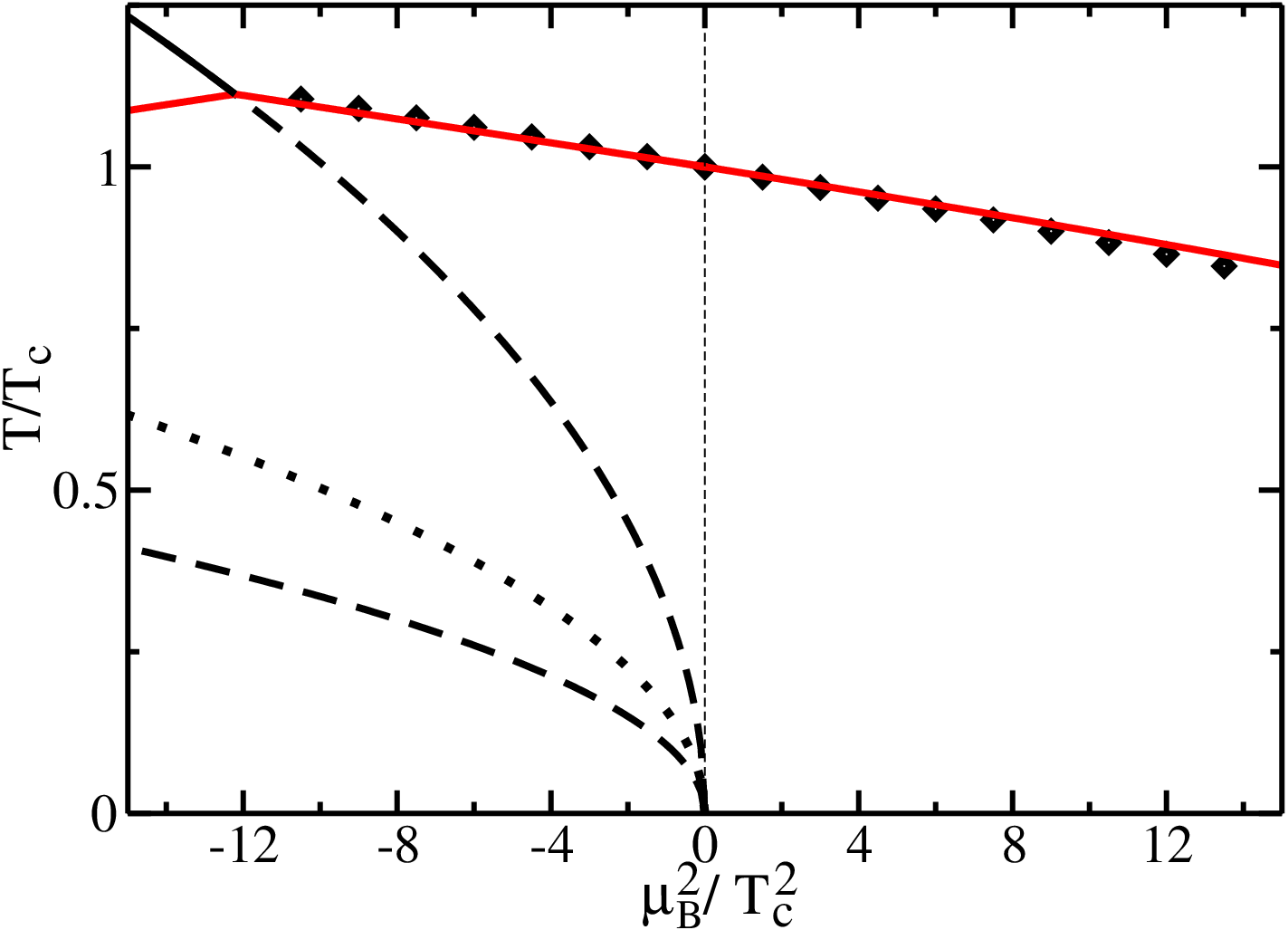}
   \hskip 3mm
   \includegraphics[scale=0.3,angle=0.]{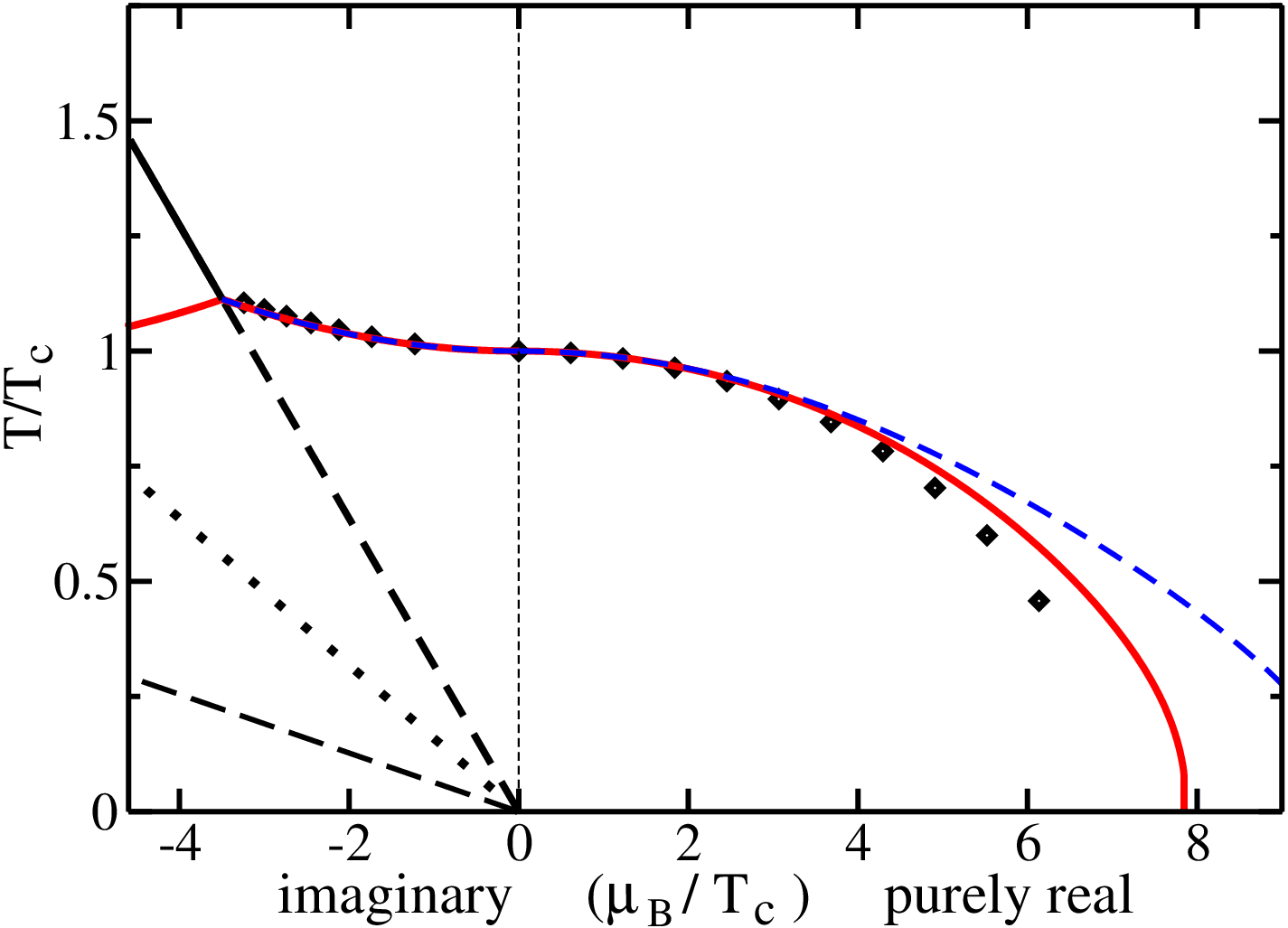}
   \caption{Phase diagram for imaginary and real 
     baryo-chemical potential. Left: $T/T_c$ 
     vs. $\mu_B^2/T_c^2$. Right: $T/T_c$ 
     vs. $\mu_B/T_c$. Details are 
     explained in the text. \label{fig.7}}
 \end{figure}

 In Fig.~\ref{fig.7}, 
 the phase diagram is exhibited in specific coordinate systems. 
 Negative 
 values $\mu_B^2\le 0$ indicate purely imaginary baryo-chemical 
 potential, whereas positive $\mu_B^2\ge 0$ indicate 
 real $\mu_B$. Diamonds represent the 
 polynomial fit~\cite{MPL2} for both, 
 imaginary chemical potential and the corresponding analytically 
 continued results. For comparison, we depict the 
 QPM characteristic (solid) curve starting at $T=T_c$ 
 as solution of Peshier's flow equation Eq.~(\ref{equ:PDE}) for imaginary 
 chemical potential 
 as well as for real $\mu_B$. The flatness of the 
 curve in the exhibited $\mu_B^2$ interval (left panel) signals 
 the dominance of the $\mu_B^2$ term in agreement with the 
 polynomial fit findings in~\cite{MPL2}. Note, however, that the QPM 
 result contains all orders of $\mu_B^2$. We emphasize that our 
 model parameters $\lambda$ and $T_s$ are adjusted to $n(T,i\mu_i)$ 
 at $T=1.1\,T_c$ and have proven above to describe at the same time 
 $n(T,i\mu_i)$ at $T=1.5, 2.5, 3.5\,T_c$. In so far, the agreement 
 of our characteristic curve emerging at $T_c$ with the transition 
 line in~\cite{MPL2} is quite satisfying. 
 
 In Fig.~\ref{fig.7} we also show the first two Roberge-Weiss 
 transition lines (fat dashed curves characterized 
 by~\cite{RW} $\mu_B^2/T_c^2=-T^2/T_c^2\pi^2(2k+1)^2$ for $k=1,2$) at 
 which thermodynamic quantities exhibit an analytic behavior at 
 small temperatures, while the Roberge-Weiss transition represents a 
 first-order phase transition (fat 
 solid section) at sufficiently large $T$. 
 In addition, the first $Z_3$ center symmetry 
 line is shown (dotted curve characterized by~\cite{RW} 
 $\mu_B^2/T_c^2=-T^2/T_c^2\pi^2(2k)^2$ for $k=1$). 
 The repeated copies of these sectors for $k\ge 2$ are not displayed 
 in Fig.~\ref{fig.7}; they reside in the left bottom edge. 
 
 Numerically, we find that the characteristic curve 
 emerging at $T=T_c$ and the first Roberge-Weiss 
 transition line cross each other at 
 $T^E/T_c=1.112$ and $(\mu_B^E)^2/T_c^2=-12.214$, whereas 
 the lattice QCD simulations~\cite{Lombardo1,MPL2} report 
 $T^E/T_c=1.095$ and $(\mu_B^E)^2/T_c^2=-11.834$. These tiny differences 
 can hardly be resolved on the scale displayed in Fig.~\ref{fig.7}. 
 For larger negative $\mu_B^2$ the 
 characteristic curve is mirrored at the Roberge-Weiss 
 transition line (see the section in the left top edge below 
 the first Roberge-Weiss transition line in the left panel 
 of Fig.~\ref{fig.7}). 
 
 In the right panel of Fig.~\ref{fig.7}, we exhibit the phase diagram 
 with the same notions as in the left panel, 
 but with a linear abscissa $\mu_B/T_c$; 
 negative values of $\mu_B/T_c$ are to be assigned to 
 purely imaginary chemical potential, while positive values 
 correspond to purely real values of $\mu_B$. 
 As the coefficients in Peshier`s flow equation Eq.~(\ref{equ:PDE}) 
 for real chemical potential obey $a_T\rightarrow 0$ for $\mu\rightarrow 0$ 
 and $a_\mu\rightarrow 0$ for $T\rightarrow 0$, the characteristic 
 curves, including the one crossing the $T$ axis at $T_c$, 
 approach the $T=0$ and $\mu_B=0$ axes perpendicularly. 
 Deviations between QPM results for real $\mu_B$ 
 and the polynomial fit become visible for 
 $\mu_B\ge 330$~MeV. In addition, we exhibit the 
 solution of Peshier's flow equation starting at $T_c$ for 
 a different set of QPM parameters (cf.~section~\ref{sec:4_1}) 
 by the dashed line. For 
 imaginary chemical potential, both results are indistinguishable, 
 whereas for real $\mu$ deviations become visible for 
 $\mu_B/T_c\ge 2$ signalling again that small deviations in the 
 imaginary chemical potential sector result in larger 
 deviations in the sector of real $\mu$. 
 This makes predictions about the 
 onset of possible deconfinement effects at small $T$ 
 and real $\mu$ difficult. 

 \subsection{Quark mass dependence \label{sec:3_3}}
 
 In this subsection, we study the influence of a different approximation 
 of the quark dispersion relation on the QPM results. In particular, 
 we concentrate on the characteristic curve emerging at $T=T_c$. 
 Apart from Eq.~(\ref{equ:dispq}), we 
 can approximate the quark dispersion relation by~\cite{Pisarski} 
 \begin{equation}
   \label{equ:dispq2}
   \omega_q^2=k^2+m_q^2+2m_qM_++2M_+^2 
 \end{equation}
 as for instance employed in~\cite{Peshier3,Bluhm-PRC} 
 with $M_+^2$ from Eq.~(\ref{equ:plafreq}). Changing 
 the approximation of the dispersion relation 
 $\omega_q$ demands a readjustment of the parameters in the 
 effective coupling $G^2(T,\mu=0)$ in Eq.~(\ref{equ:coupl}) 
 in order to appropriately describe the lattice QCD data 
 of $n/T^3$ and causes changes in Peshier's flow 
 equation (see Appendix A). 
 The QPM parameters for using Eq.~(\ref{equ:dispq2}) 
 adjusted to the continuum estimate of the $n/T^3$ data 
 at $T=1.1\,T_c$ read $T_s=0.976\,T_c$ and $\lambda = 95$ 
 implying that the divergence in $G^2(T,\mu=0)$ is located 
 at $T=0.987\,T_c$. With this new parametrization, the 
 agreement between QPM and continuum extrapolated 
 lattice QCD data 
 is nearly as perfect as observed in Fig.~\ref{fig.1} (at most 
 3\% deviations). However, it indicates 
 that Eq.~(\ref{equ:dispq}) might be somewhat more 
 suitable than Eq.~(\ref{equ:dispq2}) as quark dispersion 
 relation. The influence on the characteristic curve emerging at 
 $T=T_c$ when employing Eq.~(\ref{equ:dispq2}) instead 
 of Eq.~(\ref{equ:dispq}) is negligible for imaginary 
 chemical potential. For real $\mu_B/T_c$, the difference 
 between both parametrizations is also very tiny and approximately 
 1.5\% at small temperatures. 
 
 In addition, we can discuss the quark mass dependence 
 of the found results by performing a naive 
 chiral extrapolation $m_q\rightarrow 0$. For imaginary 
 chemical potential, quark mass effects turn out to be 
 negligible independent of the specific quark dispersion 
 relation used. For real $\mu_B/T_c$, quark mass effects 
 are also small and visible only for very small temperatures. 
 The differences between using $m_q=0.2\,T$ and $m_q=0$ are 
 less than 1\% when employing Eq.~(\ref{equ:dispq}) and at 
 most 3\% when employing Eq.~(\ref{equ:dispq2}). In both 
 cases, decreasing quark masses imply a 
 larger curvature of the characteristic curves and thus 
 a smaller critical chemical potential at $T=0$. 
 A similar minor quark mass dependence with the same trend 
 when decreasing $m_q$ was found in lattice QCD 
 simulations~\cite{lattice}. 
 
 Another sensitive measure for the quark mass dependence 
 of the net baryon density $n_B$ is the chemical potential 
 dependence of the chiral condensate $<\bar{\psi}\psi>$ which 
 are related to each other by a Maxwell 
 relation~\cite{Lombardo1,MPL}. Within the QPM, the 
 $\mu$ dependence of $<\bar{\psi}\psi>$ is fairly well 
 described and will be reported elsewhere. 
 We note that simply putting $m_q=0$ but keeping 
 the parametrization of $G^2(T,\mu=0)$ fixed modifies $n/T^3$ by less than 
 1\% for the considered range of temperatures and chemical 
 potentials. In principle, however, a general quark mass 
 dependence of the parameters $T_s$ and $\lambda$ in 
 the effective coupling would be conceivable. Due to the minor 
 effects observed, we restrict our further considerations to quark 
 dispersion relation Eq.~(\ref{equ:dispq}) in the following. 
 
 \subsection{{\boldmath$\mu$} dependence of the quasiparticle masses \label{sec:3_4}}

 In~\cite{Shuryak}, the lattice QCD data~\cite{Allton} 
 have been discussed with the goal to extract the relevant 
 excitation modes from thermodynamic bulk quantities. 
 The explicit $\mu$ dependence of the quasiparticle masses has 
 been named BKS effect. 
 In order to test the importance of the BKS effect 
 on the found results, we omitted the
 $\mu_i^2/\pi^2$ terms in the quasiparticle dispersion 
 relations 
 or flipped their signs 
 though leaving the dependence of $G^2$ on $\mu_i$ 
 unchanged. 
 In fact, neglecting simply the term $\mu_i^2/\pi^2$ in 
 Eq.~(\ref{equ:plafreq}) (or Eq.~(\ref{equ:Debmass}) below), 
 $n/T^3$ is only affected for large $\mu_i$, 
 where the attenuation of $\mu_i^2$ by $1/\pi^2$ becomes smaller and the 
 term proportional to $\mu_i^2$ 
 cannot be neglected compared to the term 
 proportional to $T^2$. This implies that for larger 
 $T$ significant effects can only be seen at sufficiently large values 
 of $\mu_i$. Note, however, that $\mu_i$ is restricted by 
 $\mu_i\le \frac \pi 3 T$. Similar effects can be observed 
 when flipping the signs in the asymptotic mass expressions. 
 Nonetheless, thermodynamic self-consistency of the QPM 
 requires in both considered cases changes in Peshier's 
 flow equation~(\ref{equ:PDE}) rendering the coefficients $b$, $a_T$ and 
 $a_{\mu_i}$ according to Maxwell's relation. In such a 
 thermodynamically self-consistent approach, we find our 
 results 
 to be indistinguishable from the QPM results 
 exhibited in Fig.~\ref{fig.1}, i.~e. found results seem to 
 be rather independent of the explicit form of the 
 $\mu_i$ dependence in the asymptotic mass expressions. 
 However, a general dependence of the asymptotic masses on 
 the chemical potential seems to be important. 
 When neglecting $\mu_i$ completely in the quasiparticle 
 dispersion relations, thermodynamic self-consistency 
 dictates also an independence of $T$ in $M_\infty$ (and 
 $m_\infty$ in Appendix A) which significantly changes the results: 
 even though the almost linear behavior 
 of $n/T^3$ for small $\mu_i$ can be reproduced, the pronounced curvature 
 for $T=1.1\,T_c$ at larger $\mu_i$ cannot be obtained under such 
 an assumption. The $\mu_i$ dependence of 
 $n/T^3$ is further discussed in Appendix B. 
   
 \subsection{Scaling properties \label{sec:3_5}}
 
\begin{figure}[b]
   \vskip 5mm
   \includegraphics[scale=0.3,angle=0.]{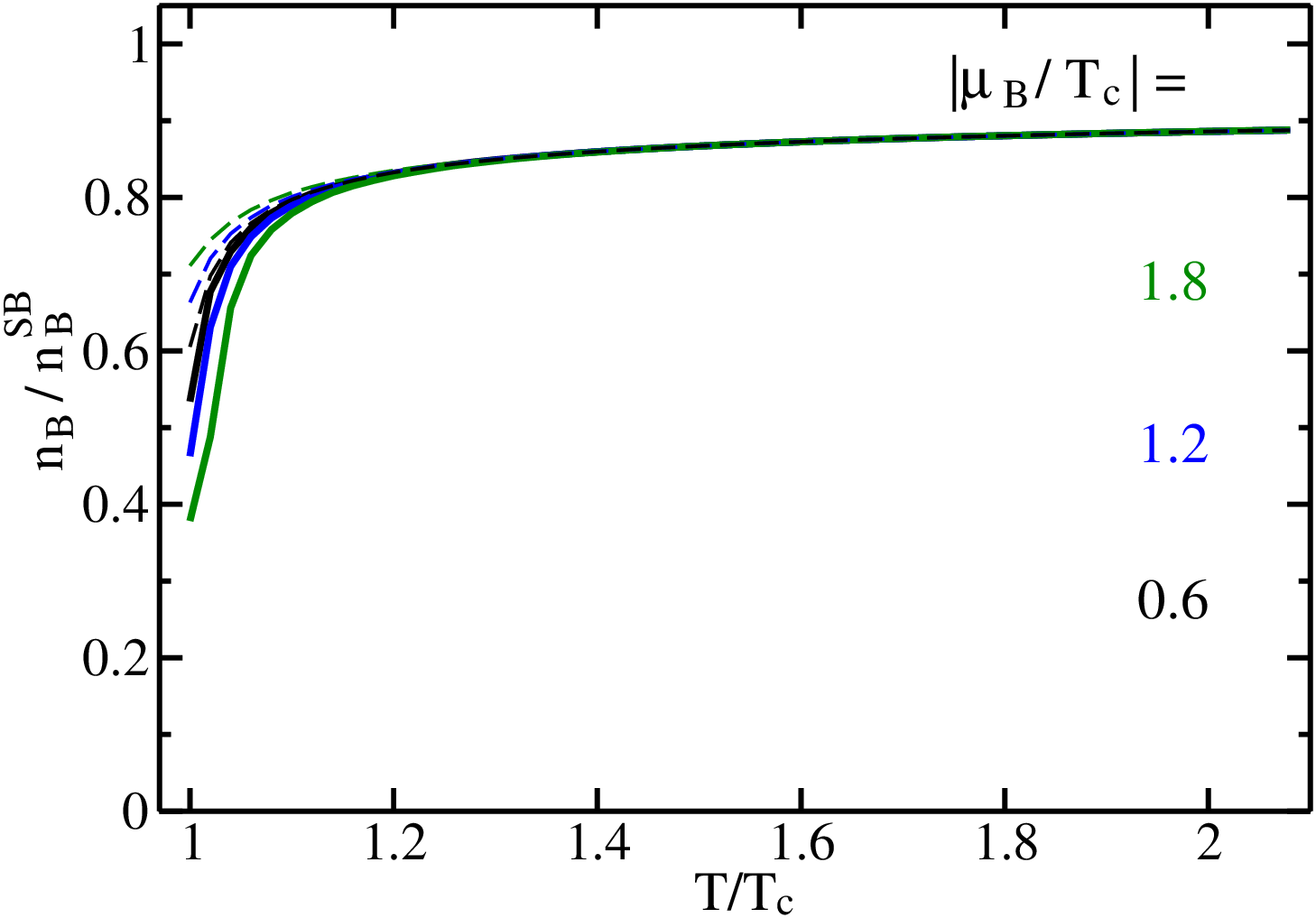}
   \includegraphics[scale=0.3,angle=0.]{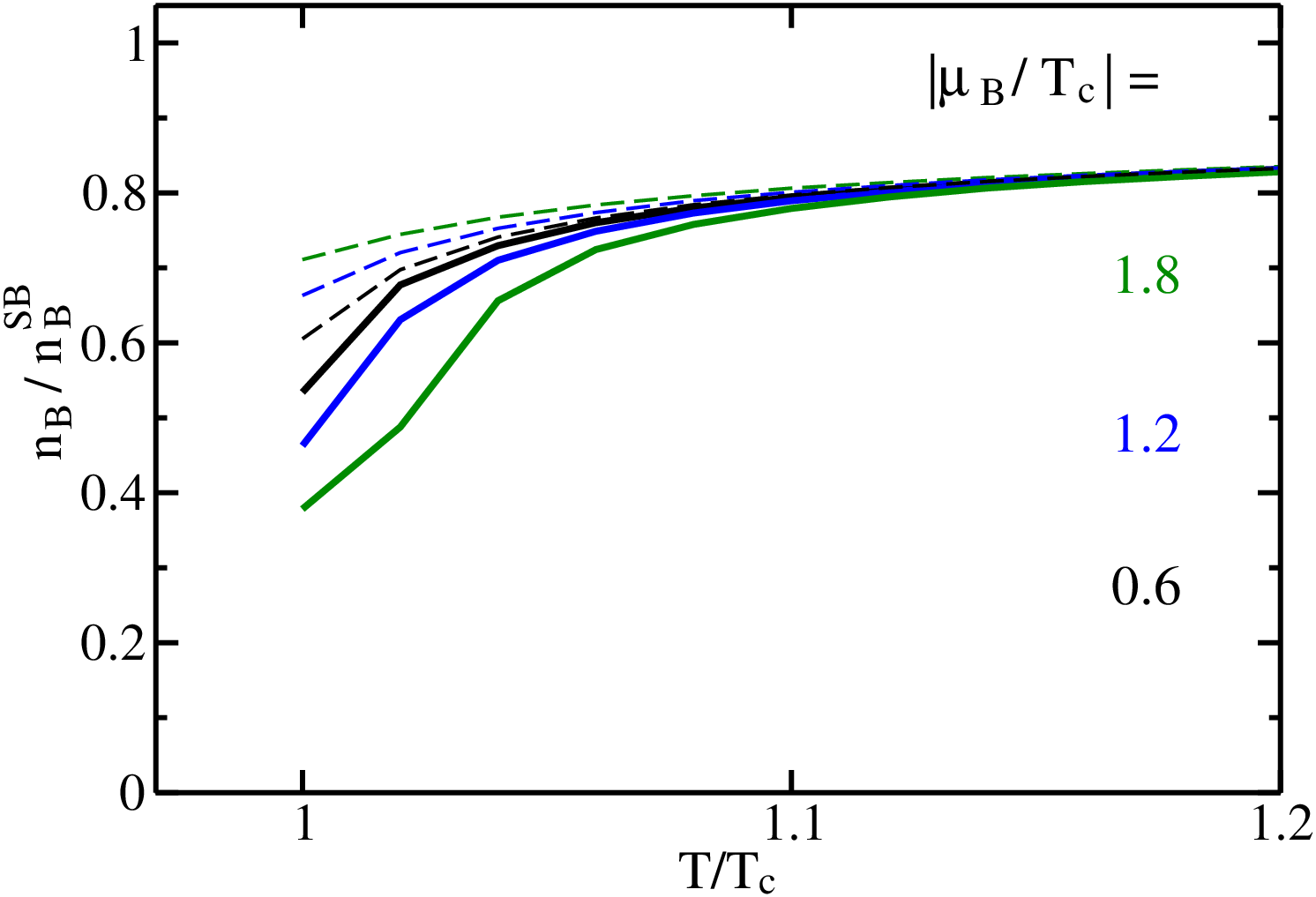}
   \caption{Left: ratio $n_B/n_B^{SB}$ as a function of 
   $T/T_c$ for different imaginary and real 
   baryo-chemical potentials. 
   Dashed curves represent 
   results for imaginary baryo-chemical 
   potential, with $|\mu_B/T_c| = 0.6, 1.2, 1.8$ 
   from bottom to top, while solid curves 
   depict corresponding results for real $\mu_B$, with 
   $|\mu_B/T_c| = 0.6, 1.2, 1.8$ in inverted order, 
   i.~e.~from top to bottom. 
   Right: zoom into the region close to $T_c$. \label{fig.B4}}
 \end{figure}

 In~\cite{Szabo,Fodor}, a scaling of the 
 ratio $\Delta p/\Delta p^{SB}$ of the excess pressure 
 in $\mu_B$ direction was 
 reported. Here, we find a similar scaling for the ratio $n_B/n_B^{SB}$ 
 as depicted in Fig.~\ref{fig.B4}, where $n_B^{SB}$ denotes the 
 Stefan-Boltzmann expression of the net baryon density. 
 When considering $n_B/n_B^{SB}$ either for real or for imaginary 
 chemical potential, in both cases, $\mu_B$ effects become 
 visible only in the vicinity of $T_c$. In fact, for the 
 baryo-chemical potentials considered in Fig.~\ref{fig.B4}, 
 the ratio $n_B/n_B^{SB}$ is found 
 to be independent of $\mu_B$ for $T\ge 1.2\,T_c$. Furthermore, 
 $n_B/n_B^{SB}=1$ is approached only asymptotically, signalling the 
 expected strong deviations from the free field behavior. Apart from 
 the observed differences in the ratio between real and imaginary chemical 
 potentials close to $T_c$, we find an interesting pattern: 
 $n_B/n_B^{SB}$ decreases with increasing 
 baryo-chemical potential for real $\mu_B$, while the ratio increases 
 in the case of imaginary chemical potential. 
 This is partly caused by differences in $n_B/T^3$ between 
 real and imaginary chemical 
 potential which become smaller for increasing temperature 
 (cf. Fig.~\ref{fig.4}). But it is also related to 
 different signs in $n_B^{SB}$ for real or imaginary chemical 
 potential. 
 
 To be specific, in the case of real chemical potential, 
 $n_B$ can be expanded into a 
 Taylor series in powers of $\mu_B$ with expansion coefficients~\cite{Allton} 
 $c_k(T)=\frac{1}{k!} \left.\frac{\partial^k (p(T,\mu)/T^4)}{\partial (\mu/T)^k}
 \right|_{\mu=0}$. 
 The Stefan-Boltzmann expression for the net baryon density reads 
 \begin{equation}
   n_B^{SB}(T,\mu_B)=\frac{N_f}{3}\frac{\mu_B}{3}T^2 + 
   \frac{N_f}{3\pi^2} \left(\frac{\mu_B}{3}\right)^3 \,.
 \end{equation}
 Even though this expression for $n_B^{SB}$ is correct only for a massless 
 ideal gas, while $n_B$ entering the ratio is evaluated for 
 $m_q=0.2\,T$, quark mass effects 
 can safely be neglected (as discussed in section~\ref{sec:3_3}). 
 The ratio $n_B/n_B^{SB}$ reads 
 \begin{equation}
   \frac{n_B}{n_B^{SB}} \approx \frac{2c_2}{N_f} + 
   \frac{2}{N_f}\left(\frac{\mu_B}{3T}\right)^2 \left[2c_4-
   \frac{c_2}{\pi^2}\right] + \mathcal{O}(\mu_B^4) \,.
 \end{equation}
 In the limit $\mu_B\rightarrow 0$, the ratio approaches 
 $2c_2/N_f=\frac 14 \chi (T,\mu=0)/T^2$ for $N_f=4$. 
 For small $\mu_B/(3T)$, i.~e.~for small $\mu_B$ or large $T$, 
 $\mu_B$ effects become small, thus explaining the 
 observed scaling. Furthermore, as $2c_4-c_2/\pi^2 > 0$ for all 
 temperatures $T\ge T_c$ and remains approximately constant 
 for $T\ge 1.2\,T_c$ (cf.~section~\ref{sec:4_2}), a fixed 
 ratio $n_B/n_B^{SB}$ requires increasing temperatures $T$ for 
 increasing $\mu_B$, explaining the 
 observed ordering in Fig.~\ref{fig.B4}. Close 
 to $T_c$, deviations between exact results and the Taylor 
 series expansion of $n_B$ become larger with increasing 
 $\mu_B$, such that the arguments presented here do not 
 apply. 

 In the case of imaginary chemical potential, $n_B(T,i\mu_i)$ 
 from Eq.~(\ref{equ:dens1}) can be evaluated for small $\mu_i$ by expanding the 
 trigonometric functions in powers of $\mu_i/T$ yielding also a 
 Taylor series expansion similar to the one in the sector 
 of real chemical potential. Within this approach, we find 
 \begin{equation}
   n_B(T,i\mu_i) = i\left(\frac 23 \tilde{c}_2\mu_i T^2 - 
   \frac 43 \tilde{c}_4 \mu_i^3 + ... \right) \,,
 \end{equation}
 where 
 \begin{equation}
   \tilde{c}_k(T) = 
   \frac{1}{k! i^k} \left.\frac{\partial^k (p(T,i\mu_i)/T^4)}{\partial (\mu_i/T)^k}
   \right|_{\mu_i=0} \equiv c_k(T)\,.
 \end{equation}
 Note that both, $\tilde{c}_k$ and $c_k$ are real and 
 $\tilde{c}_k,c_k=0$ for odd $k$. 
 The Stefan-Boltzmann result for imaginary chemical potential 
 reads 
 \begin{equation}
   n_B^{SB}(T,i\mu_i) = i\left(\frac{N_f}{3}\mu_i T^2 - \frac{N_f}{3\pi^2} \mu_i^3
   \right) 
 \end{equation}
 and the ratio follows as $\frac{n_B}{n_B^{SB}} \approx \frac{2\tilde{c}_2}{N_f} - 
   \frac{2}{N_f}\left(\frac{\mu_B}{3T}\right)^2 \left[
   2\tilde{c}_4 - \frac{\tilde{c}_2}{\pi^2} \right] + \mathcal{O}(\mu_B^4)$. 
 Similar to the considerations for real chemical potential, we observe 
 a scaling with $\mu_B/(3T)$ and in the limit $\mu_B\rightarrow 0$, 
 $n_B/n_B^{SB}\rightarrow 2c_2/N_f=\frac 14 \chi (T,\mu=0)/T^2$ 
 for $N_f=4$. 
 For imaginary chemical potential, however, the sign of the term proportional 
 to $\mu_B^2$ is flipped, explaining the different ordering observed 
 in Fig.~\ref{fig.B4}, i.~e. at fixed $T$, $n_B/n_B^{SB}$ becomes larger 
 with increasing $\mu_B$. 
 
 \section{Comparison with lattice QCD data at 
 {\boldmath$\mu=0$} \label{sec:4}}

 \subsection{Pressure \label{sec:4_1}}
 
 Via the QPM, we have access to both, 
 real and imaginary chemical potentials. Thus, we can 
 compare our results based on the lattice QCD data of~\cite{Lombardo1,MPL} 
 with other lattice QCD calculations. 
 In~\cite{Engels}, a similar lattice setup for 
 calculating the pressure at $\mu=0$ for 
 $N_f=4$ degenerate quark flavors with $m_q=0.2\,T$ on a lattice with 
 $N_\tau=4$ and $N_\sigma=16$ was considered, 
 though employing an improved lattice action. 
 These lattice QCD data~\cite{Engels} require also a proper 
 continuum extrapolation. We apply a similar strategy as 
 in section~\ref{sec:3_1} but now for the pressure, because 
 its Stefan-Boltzmann limit is given in~\cite{Engels} for $N_\tau=4$, 
 and find $d_{lat}^{(p)}=0.839$ as continuum extrapolation factor. 
 The difference between $d_{lat}^{(p)}$ and $d_{lat}^{(n)}$ in 
 section~\ref{sec:3} is maybe a consequence of the different 
 lattice actions used in the simulations~\cite{Engels} 
 and~\cite{Lombardo1,MPL} resulting in different 
 cut-off effects on the data. 
 
\begin{figure}[t]
   \includegraphics[scale=0.35,angle=0.]{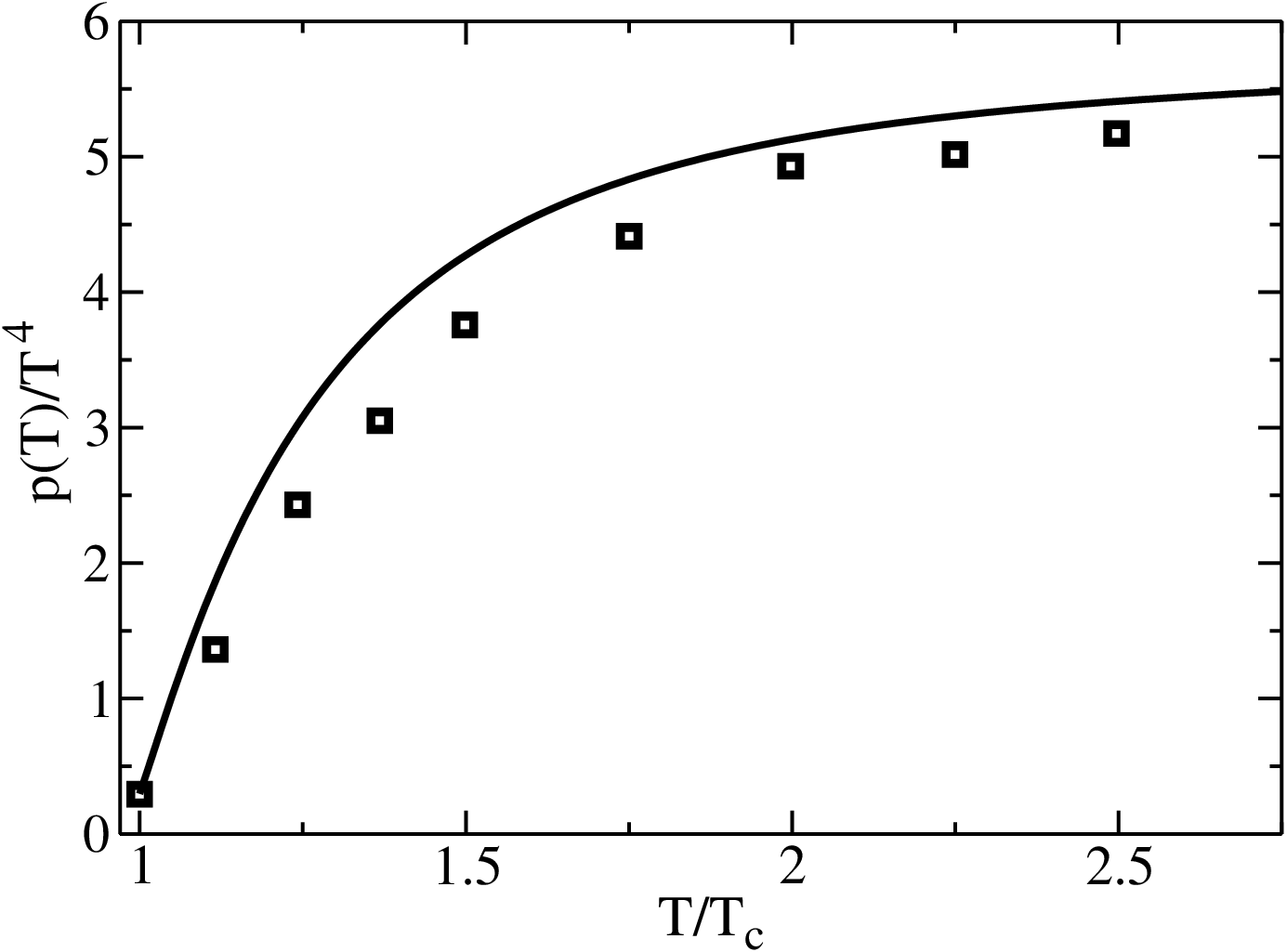}
   \caption{Comparison of the QPM (solid curve, using parameters 
     as in section~\ref{sec:3_1}) for the scaled pressure 
     $p/T^4$ as a function of $T/T_c$ at 
     $\mu=0$ with the continuum estimate of the lattice 
     QCD data~\cite{Engels} (squares).
     \label{fig.5}}
 \end{figure}

 In Fig.~\ref{fig.5}, 
 the continuum estimated lattice QCD data~\cite{Engels} (squares) for 
 $p/T^4$ as a function of $T/T_c$ at $\mu=0$ are 
 compared with the QPM using the 
 parameters $\lambda$ and $T_s$ in $G^2(T,\mu=0)$ 
 from section~\ref{sec:3_1} adjusted to $n/T^3$ for 
 imaginary chemical potential. 
 The integration constant $B(T_c)$ adjusting the QPM 
 value of $p/T^4$ at $\mu=0$ and $T=T_c$ to lattice 
 QCD reads $B(T_c)=2.56\,T_c^4$ using again 
 $T_c=163$~MeV. As evident from Fig.~\ref{fig.5}, the 
 general trend of $p/T^4$ and the behavior at large $T$ 
 is reproduced. Nevertheless, $p/T^4$ shows deviations of up to 
 20\% in the intermediate temperature region $T\sim 1.5\,T_c$. 
 The same deviation pattern was already discussed 
 in~\cite{Bluhm-PLB,Szabo,Bluhm-PRC}. In fact, it seems to be a 
 general feature, that fits to lattice QCD data in 
 the sector of zero (non-zero) chemical potential 
 underestimate (overestimate) the according results 
 in the sector of non-zero (zero) chemical potential. 
 
\begin{figure}[b]
   \vskip 5mm
   \includegraphics[scale=0.3,angle=0.]{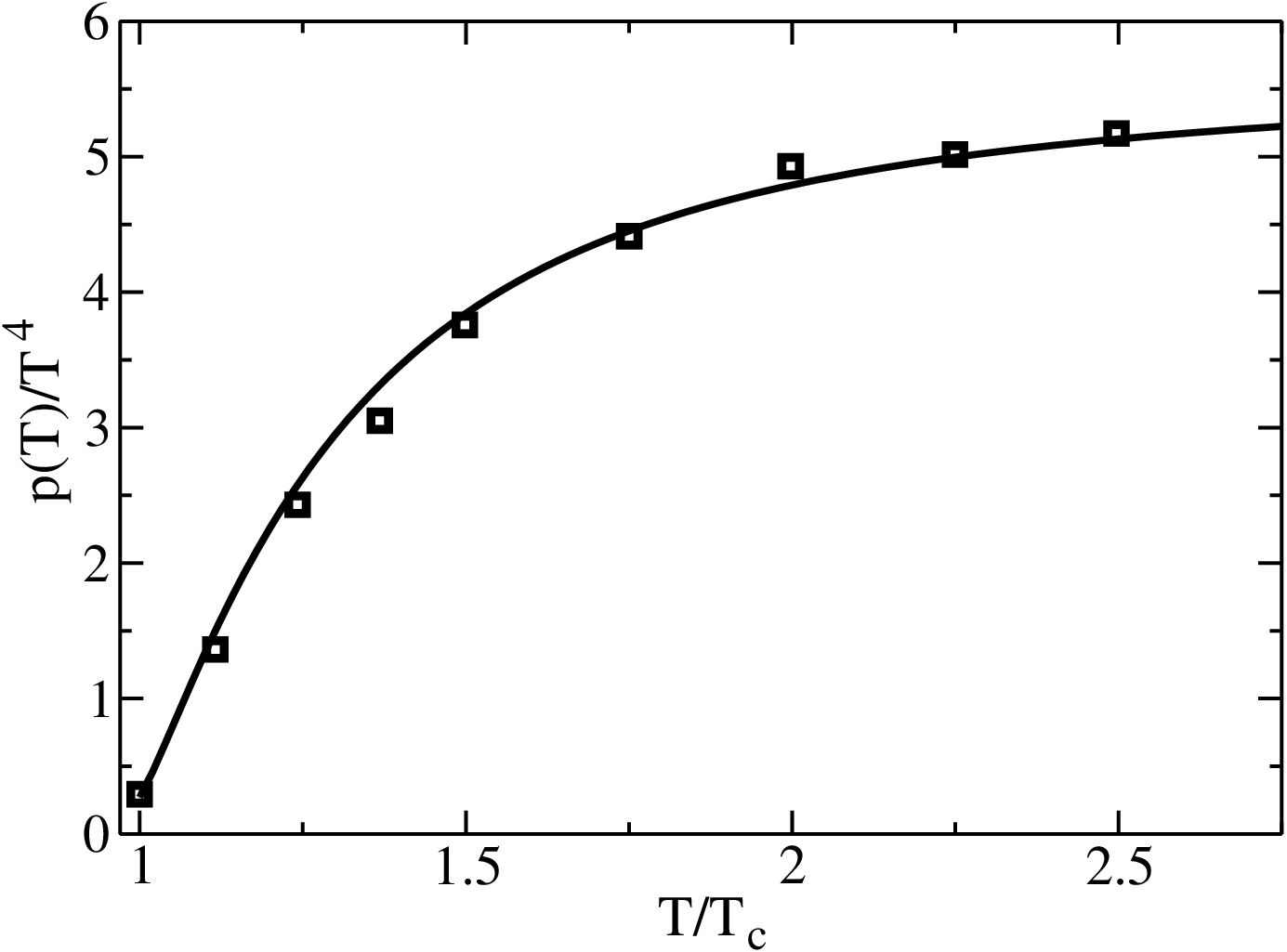}
   \hskip 3mm
   \includegraphics[scale=0.3,angle=0.]{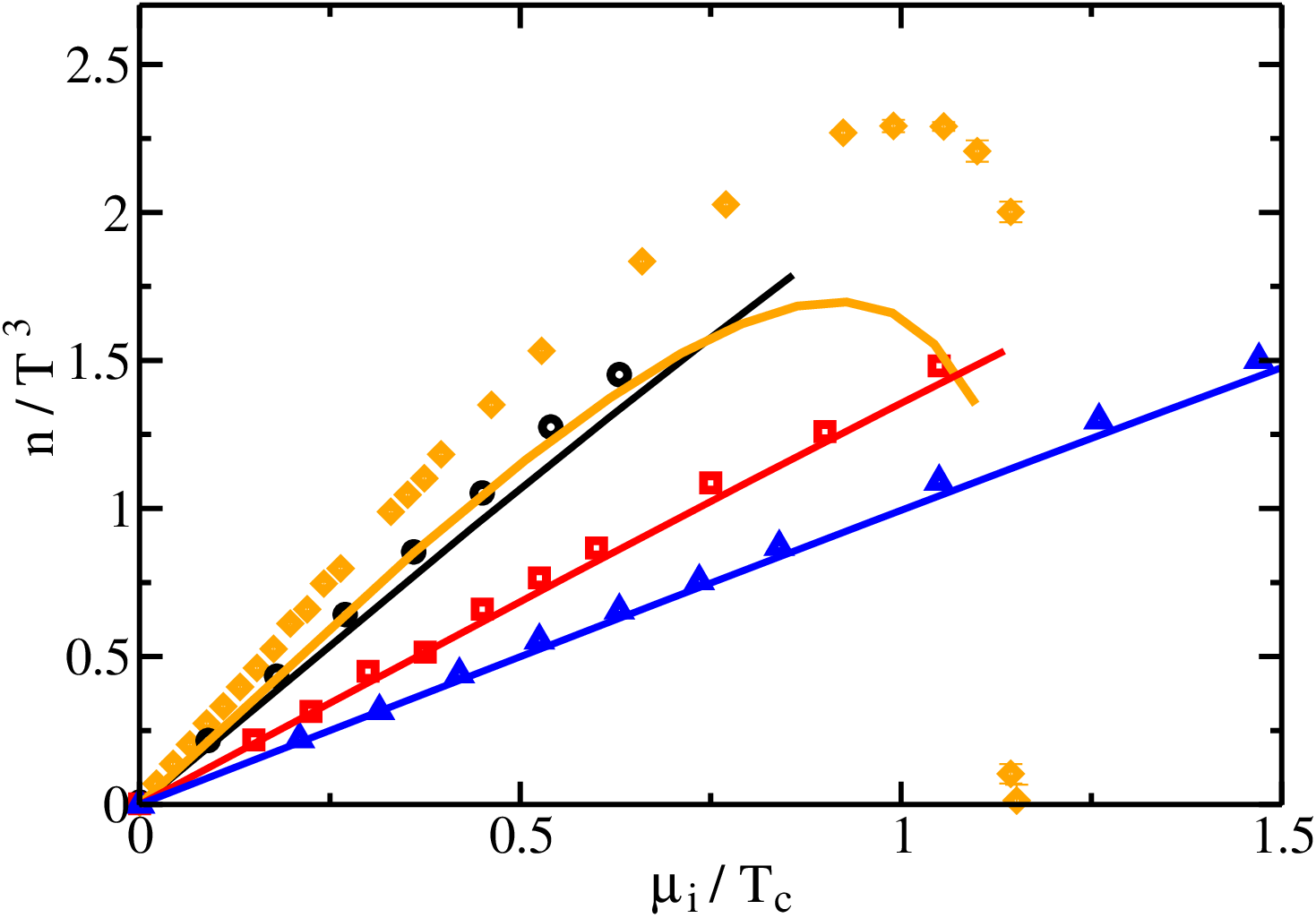}   
   \caption{Left: comparison of the QPM (solid curve) 
     for $p/T^4$ as a 
     function of $T/T_c$ at $\mu=0$ employing readjusted 
     parameters (see text) with the continuum extrapolated 
     lattice QCD data (squares) as exhibited in 
     Fig.~\ref{fig.5}. 
     Right: comparison of the QPM (solid curves, using the 
     readjusted parameters) for $n/T^3$ as a function of 
     $\mu_i/T_c$ for different $T$ with the continuum 
     extrapolated lattice QCD data (symbols) as 
     exhibited in Fig.~\ref{fig.1}. \label{fig.6}}
 \end{figure}

 Considering, instead, an independent adjustment of the QPM 
 parameters to $p/T^4$ at $\mu=0$, the 
 comparison of the QPM with the continuum estimate of the 
 lattice QCD data is 
 exhibited in Fig.~\ref{fig.6} (left panel). 
 We find an impressive agreement when adjusting 
 $T_s=0.91\,T_c$, $\lambda=16$ and $B(T_c)=1.25\,T_c^4$ 
 with $T_c=163$~MeV. With this new parametrization, 
 we evaluate the net quark number density for 
 imaginary chemical potential. The results (solid 
 curves) are shown in the right panel of 
 Fig.~\ref{fig.6}. At constant $\mu_i/T_c$, we find 
 increasing deviations for decreasing temperatures, 
 in particular close to the Roberge-Weiss 
 transition, even though the change in slope 
 close to $\mu_c$ is qualitatively still reproduced. 
 Furthermore, using this 
 set of parameters results in the dashed curve in the right panel of 
 Fig.~\ref{fig.7} as characteristic curve emerging at $T=T_c$ 
 for real $\mu_B$. 
 With respect to the observed deviations at 
 $T=1.1\,T_c$ between the lattice QCD data 
 from~\cite{Lombardo1,MPL} for $n/T^3$ and the QPM with parameters 
 adjusted to $p/T^4$ (see right panel of Fig.~\ref{fig.6}), it is 
 surprising that this characteristic curve agrees so well with 
 the one considered in section~\ref{sec:3_2}. 

 \subsection{Taylor expansion coefficients \label{sec:4_2}}
 
\begin{figure}[t]
   \includegraphics[scale=0.35,angle=0.]{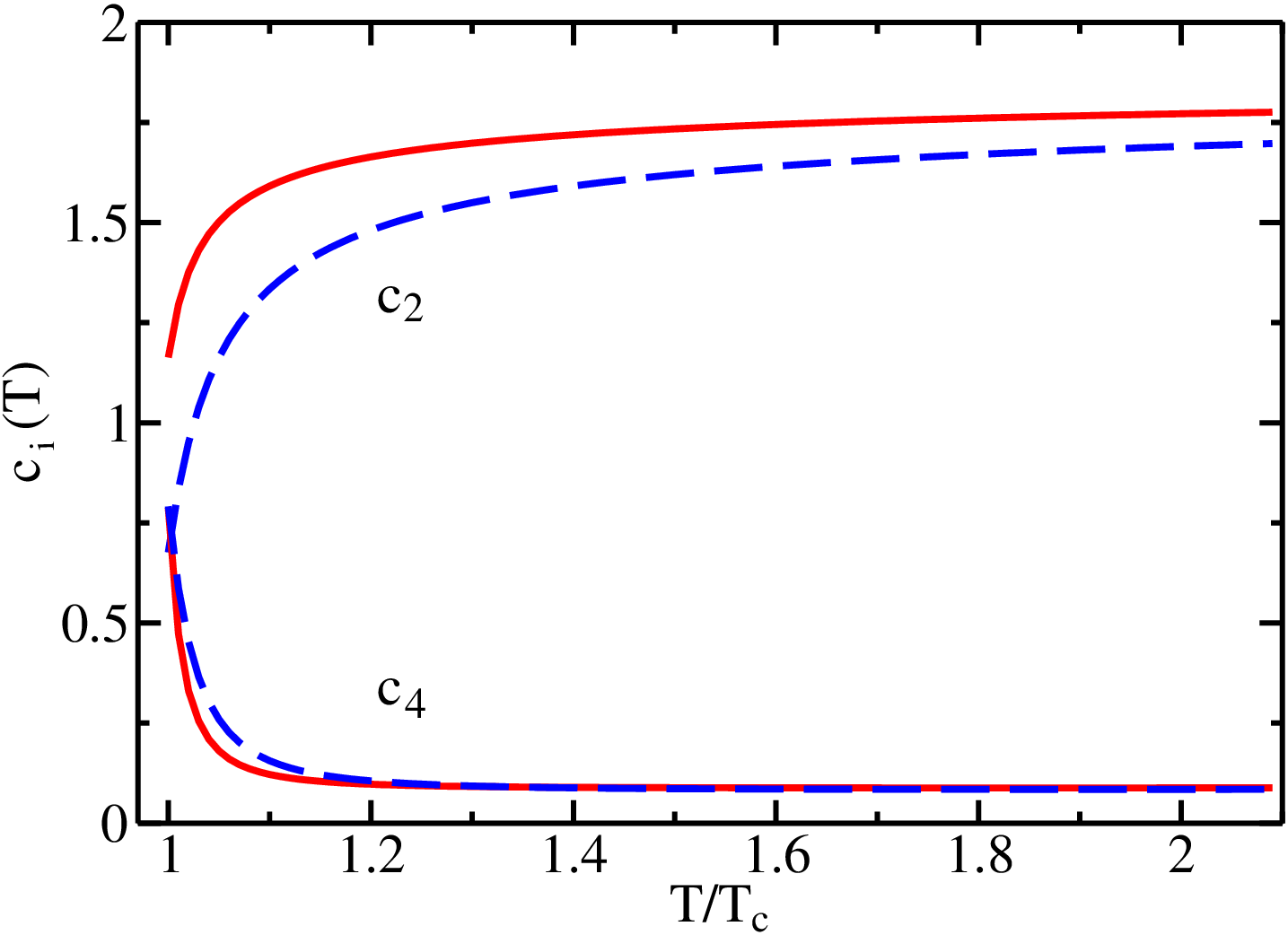}
   \caption{Taylor series expansion coefficients 
   $c_2(T)$ (upper curves) and $c_4(T)$ 
   (lower curves) as function of $T/T_c$ for $N_f=4$ 
   employing the different parametrizations from Fig.~\ref{fig.1} 
   (solid curves) and Fig.~\ref{fig.6} (dashed curves). \label{fig.B2}}
 \end{figure}

 Having the QPM parametrizations employed in 
 Figs.~\ref{fig.1} and~\ref{fig.6} at hand, we can discuss 
 their influence on the Taylor series expansion coefficients 
 $c_k(T)$ defined in section~\ref{sec:3_5} for $N_f=4$ similar to studies 
 for $N_f=2$ in~\cite{Bluhm-PLB}. In Fig.~\ref{fig.B2}, 
 we exhibit $c_2(T)$ and $c_4(T)$: $c_2(T)=\frac 12 \chi (T,\mu=0)/T^2$ 
 shows some deviations between both parametrizations whereas $c_4(T)$ 
 agrees fairly well with visible deviations only in the 
 very vicinity of $T_c$. In addition, we observe 
 that smaller $c_0(T)\equiv p(T,\mu=0)/T^4$ 
 (cf.~Fig.~\ref{fig.5} and the left panel of Fig.~\ref{fig.6}) 
 implies smaller $c_2(T)$ as already pointed out in~\cite{Bluhm-PRC}. 
 As already mentioned in section~\ref{sec:3_1}, $n/T^3$ depicted as 
 a function of $\mu_i/T$ shows almost no temperature dependence 
 for $T\ge 1.5\,T_c$. This is mainly due to the fact that $c_2(T)$ 
 (upper solid curve in Fig.~\ref{fig.B2}) exhibits also a rather 
 negligible temperature dependence for larger $T$. Furthermore, 
 $c_4(T)$ is sizeable only close to $T_c$ and approaches its 
 Stefan-Boltzmann limit $1/\pi^2$ for $T\ge 1.2\,T_c$. Considering 
 $n_B/T^3$ in terms of a Taylor series expansion up to order 
 $\mathcal{O}(\mu^3)$, the results for real and imaginary chemical 
 potential differ only in the sign of the cubic term which is 
 $\propto c_4 \mu^3$. Thus, the net baryon density evaluated for 
 real or imaginary chemical potential deviates only for larger 
 chemical potentials and close to $T_c$ as evident from 
 Fig.~\ref{fig.4}. $c_6(T)$ (not exhibited) 
 deviates significantly from 
 zero only for temperatures very close to $T_c$ but can become 
 of the same order of magnitude as $c_4(T)$ at $T=T_c$. 
 
 Finally, we mention that 
 \begin{figure}[b]
   \vskip 5mm
   \includegraphics[scale=0.3,angle=0.]{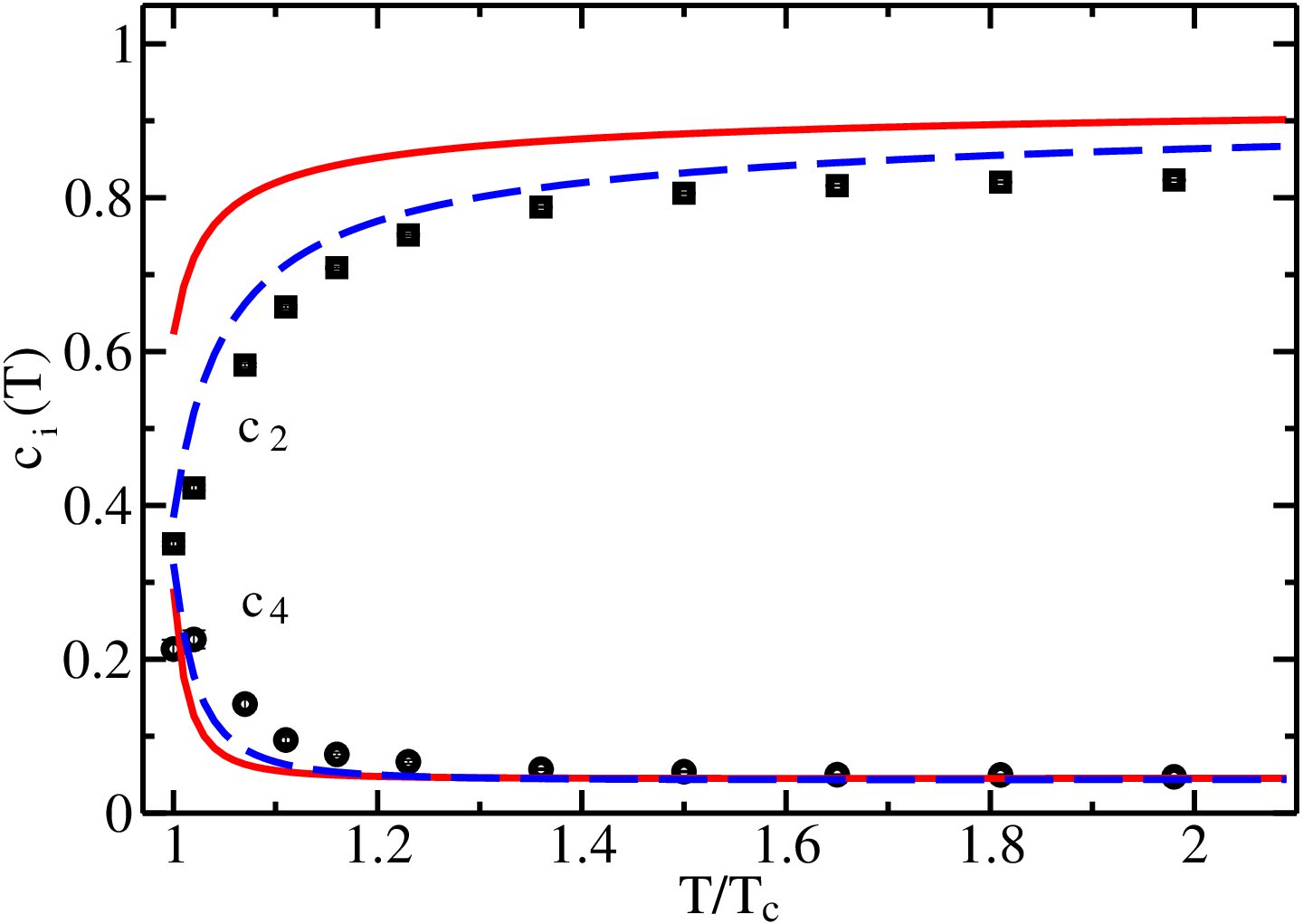} 
   \hskip 2mm
   \includegraphics[scale=0.3,angle=0.]{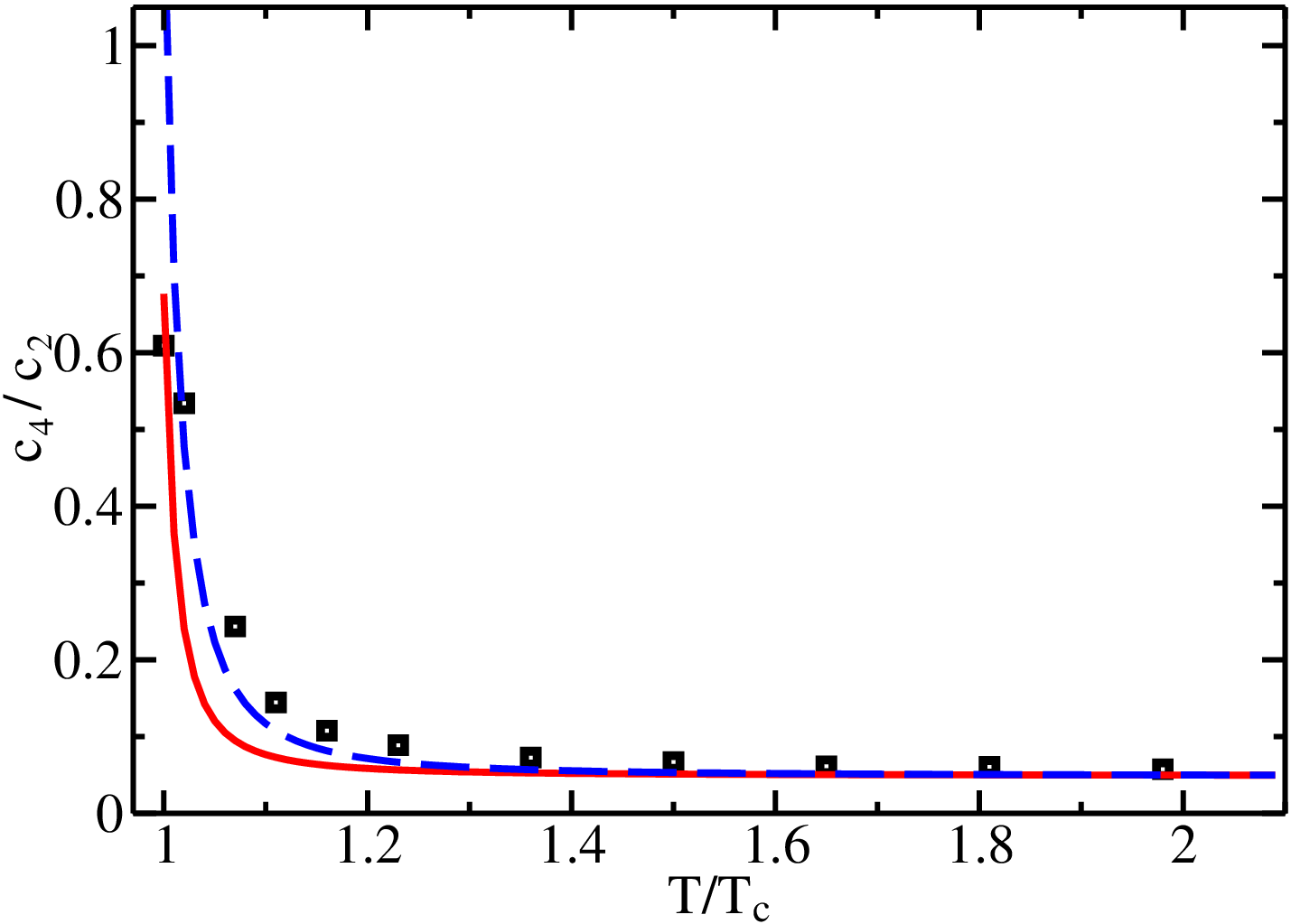}
   \caption{Left: comparison of lattice QCD data~\cite{Allton} for $N_f=2$ 
     for $c_2(T)$ (squares) and $c_4(T)$ (circles) as a function 
     of $T/T_c$ with the QPM for $N_f=2$. Solid curves 
     represent results applying QPM parameters as in 
     Fig.~\ref{fig.1} adjusted to $n/T^3$ while dashed 
     curves represent results applying the parametrization 
     of Fig.~\ref{fig.6} adjusted to $p/T^4$ at $\mu=0$. 
     Right: ratio $c_4/c_2$ as a function of 
     $T/T_c$ for both parametrizations. \label{fig.B3}}
 \end{figure} 
 the parametrization in section~\ref{sec:4_1}, optimized 
 for reproducing the $N_f=4$ lattice QCD data of~\cite{Engels}, 
 can also be used to describe the Taylor coefficients $c_{2,4}(T)$ 
 from lattice QCD~\cite{Allton} for $N_f=2$ as exhibited in the left 
 panel of Fig.~\ref{fig.B3} (dashed curves). In doing so, we keep 
 $\lambda$ and $T_s$ adjusted to $N_f=4$ lattice QCD data fixed 
 and change merely from $N_f=4$ to $N_f=2$ in the 
 thermodynamic expressions of the QPM. In fact, we find 
 deviations of about 10\% close to $T_c$ and less than 5\% 
 for $T\ge 1.2\,T_c$ between QPM and~\cite{Allton} for $c_2(T)$.  
 While this coincidence might be 
 accidental, one could also argue that the quasiparticle model catches correctly 
 the flavor dependence. In contrast, employing the parametrization 
 from section~\ref{sec:3_1} gives a pattern resembling 
 Fig.~\ref{fig.B2}. Even though deviations between both parametrizations 
 are obvious, the ratio $c_4/c_2$ is rather insensitive with respect 
 to the employed parametrization for $T\ge 1.2\,T_c$ approaching 
 $1/(2\pi^2)$ as shown in the right panel of Fig.~\ref{fig.B3}. 
    
 \section{Summary and Discussion \label{sec:5}}
 
 In summary we extend our effective quasiparticle model and compare it with
 lattice QCD data for purely imaginary chemical potential. Despite the fact
 that our phenomenological model does not exhibit the Roberge-Weiss
 periodicity of full QCD, it is able to describe the available lattice QCD 
 data~\cite{Lombardo1,MPL,MPL2} 
 impressively well. In particular, the drastic change in slope 
 of $n/T^3$ close to the critical chemical potential $\mu_c/T=\pi /3$ 
 of the Roberge-Weiss transition can be described. This is entirely 
 due to the BKS effect~\cite{Shuryak}, i.~e.~a consequence of chemical 
 potential dependent quasiparticle masses. 
 A thermodynamically consistent investigation 
 of the importance of the $\mu_i$ dependence in the 
 quasiparticles' asymptotic masses 
 shows that the found results are independent 
 of the chosen explicit form of the $\mu_i$ dependence. 
 Nonetheless, the pronounced 
 structures cannot be reproduced when the quasiparticle 
 masses would be completely independent of $\mu_i$. In
 this respect, the $\mu_i$ dependence implemented in the model
 is confirmed. Another evidence is the comparison of 
 the QPM result for the characteristic curve emerging 
 at $T=T_c$ with the phase transition line evaluated in 
 lattice QCD simulations~\cite{MPL2}. 
 For the Roberge-Weiss transition, we find critical 
 values of temperature and baryo-chemical potential close 
 to the ones given in~\cite{MPL2}. 
 The successful comparison points to the correctness 
 of Peshier's flow equation as a tool for transporting information from 
 $\mu = 0$ to non-zero $\mu$ which is of particular importance for 
 the knowledge of the equation of state at larger baryon densities 
 relevant for CERN-SPS and upcoming FAIR. 
 
 With the found QPM parametrization describing lattice QCD data 
 in the sector of purely imaginary chemical potential at hand, 
 we can also compare with an 
 independent set of lattice QCD 
 data~\cite{Engels} obtained at $\mu=0$. 
 We find some deviations for the pressure 
 in the intermediate temperature region which might account 
 for the different lattice actions used in the calculations 
 but could also signal to some extend a disagreement of results 
 obtained at $\mu =0$ and non-zero $\mu$, as already 
 discussed in~\cite{Bluhm-PLB,Bluhm-PRC}. In this context we emphasize 
 that the comparison of thermodynamic models with lattice 
 QCD data is hampered by the lacking systematic continuum extrapolation 
 of the latter. 
 
 Ab initio it is not clear whether the
 assumed quasiparticle excitations represent 
 the proper description of QCD thermodynamics also in the
 region close to $T_c$. The success of the present comparison lends some
 credibility into the picture of quasiquark excitations with a mass
 gap. This in in line with findings in \cite{Karsch_Kitazawa}, where
 also a striking deviation from the pertubative excitation pattern 
 close to $T_c$ 
 has been found. Nevertheless, it would be premature to claim that
 the strongly coupled hot quark-gluon medium is entirely described by
 the presently used quasiparticles. For instance, excitation modes
 like plasminos and longitudinal gluons are not included in the model. 
 Furthermore, finite width effects of the quasiparticles and Landau damping are
 neglected. One should keep in mind that thermodynamic bulk
 properties are sensitive essentially to excitations with hard
 momenta, i.e. $k \sim T, \mu$. There may be a variety of soft and
 ultra-hard excitations rendering the picture of the strongly coupled
 quark gluon plasma into a much more involved scenario, 
 in line with the complexity of QCD.
 
 Our model is far from being an ab initio calculation as attempted in
 \cite{Blaizot}. But, in particular, the flexibility of the
 introduced effective coupling $G^2$ allows for curing possible deficits in the
 dynamical degrees of freedom. Apart from that, the model is highly
 non-perturbative as it can be formulated in terms of an infinite series of powers in
 the coupling, though, making contact with 
 perturbation theory, as the first terms coincide with perturbative
 QCD and asymptotically $G^2$ approaches the running QCD coupling. 
 
 Finally, we remind the reader that we consider here a fairly special case 
 of four degenerate quark flavors. Despite of the known sensitivity of 
 particular features of QCD on the flavor content, some scaling properties 
 of thermodynamic bulk properties may be useful for an orientation 
 in thermodynamic state space. 
 
 \subsection*{Acknowledgements}
 
 The authors gratefully acknowledge stimulating and 
 enlighting discussions with M.~P.~Lombardo,
 as we also thank for supplying lattice QCD data in an early stage 
 of these investigations. Furthermore, we 
 thank P. de Forcrand for interesting conversations.
 The work is supported by BMBF 06DR136, GSI-FE and EU-I3HP. 
 
 \section*{Appendix A - Flow equation for imaginary chemical potential}
 
 The QPM pressure for imaginary chemical potential reads 
 \begin{equation}
   p(T,i\mu_i) = \sum_{a=q,g} p_a(T,i\mu_i) - B(T,i\mu_i) 
 \end{equation}
 with partial pressures 
 \begin{eqnarray}
   \label{e:ipq}
   p_q(T,i\mu_i) & = & \frac{d_q}{2\pi^2} T \int_0^\infty dk k^2 
   \left(\ln \left[1+e^{(i\mu_i-\omega_q)/T}\right]
    + \ln \left[1+e^{(-i\mu_i-\omega_q)/T}\right]\right) ,\\
   \label{e:ipg}
   p_g(T,i\mu_i) & = & - \frac{d_g}{\pi^2} T \int_0^\infty dk k^2 
   \ln \left[1-e^{-\omega_g/T}\right] 
 \end{eqnarray}
 for quarks and gluons, respectively, where $d_q=2N_cN_f$ and 
 $d_g=N_c^2-1$. The quasiparticle dispersion relations read 
 $\omega_q^2=k^2+m_q^2+2M_+^2 \equiv k^2+M_\infty^2$ with $M_+^2$ given in 
 Eq.~(\ref{equ:plafreq}) and $\omega_g^2=k^2+m_\infty^2$ with 
 asymptotic mass 
 \begin{equation}
 \label{equ:Debmass}
   m_\infty^2 = \frac{1}{12} \left( [2N_c + N_f] T^2 - 
   \frac{N_c}{\pi^2} N_f\mu_i^2 \right) G^2(T, i\mu_i) \,.
 \end{equation}
 Assuming that all $T$ and $\mu_i$ dependence of the 
 function $B$ is encoded in the asymptotic mass 
 expressions $M_\infty$ and $m_\infty$, thermodynamic consistency 
 is fulfilled from the stationarity conditions 
 $\partial p/\partial M_\infty^2 = \partial B/\partial M_\infty^2$ 
 and 
 $\partial p/\partial m_\infty^2 = \partial B/\partial m_\infty^2$~\cite{Gorenstein}
 such that entropy density $s$ and net quark number density $n$ 
 are obtained from standard thermodynamic relations. The purely 
 real result for $s=s_q+s_g$ reads 
 \begin{eqnarray}
   s_q(T,i\mu_i) & = & \frac{d_q}{2\pi^2T}\int_0^\infty dk k^2 
   \left(\frac{\frac{4}{3}k^2+M_\infty^2}{\omega_q}
   \left[f_q^++f_q^-\right] - i\mu_i\left[f_q^+-f_q^-\right]\right) \,,\\
   s_g(T,i\mu_i) & = & \frac{d_g}{\pi^2T}\int_0^\infty dk k^2 
   \frac{\frac{4}{3}k^2+m_\infty^2}{\omega_g}
   \frac{1}{e^{\omega_g/T}-1} \,,
 \end{eqnarray}
 where $f_q^\pm = (e^{(\omega_q\mp i\mu_i)/T}+1)^{-1}$ and $n$ 
 is given in Eq.~(\ref{equ:dens1}). The quasi-linear partial 
 differential equation Eq.~(\ref{equ:PDE}) to be solved for 
 $G^2(T,i\mu_i)$ follows from Maxwell's relation 
 \begin{equation}
   \frac{\partial s}{\partial (i\mu_i)} = 
   \frac{\partial^2 p}{\partial (i\mu_i)\partial T} = 
   \frac{\partial^2 p}{\partial T\partial (i\mu_i)} = 
   \frac{\partial n}{\partial T} \,,
 \end{equation}
 where the explicit derivative terms cancel each other leaving 
 \begin{equation}
  \label{e:Maxwell1}
  \frac{\partial n}{\partial M_\infty^2}\frac{\partial M_\infty^2}{\partial T} = 
  \frac{\partial s_q}{\partial M_\infty^2}\frac{\partial M_\infty^2}{\partial (i\mu_i)} + 
  \frac{\partial s_g}{\partial m_\infty^2}\frac{\partial m_\infty^2}{\partial (i\mu_i)}\,.
 \end{equation}
 Omitting the overall factor of $i$ in 
 $\partial n/\partial M_\infty^2$, 
 $\partial M_\infty^2/\partial (i\mu_i)$ and 
 $\partial m_\infty^2/\partial (i\mu_i)$, 
 the coefficients of Eq.~(\ref{equ:PDE}) read 
 \begin{eqnarray}
  \label{e:PDEb}
  b & = & \left(\frac{C_f}{2} T G^2 + 2m_qa\right){\rm I_1} - 
  \frac{N_cN_f}{6\pi^2} \mu_i G^2 {\rm I_2} - 
  \frac{C_f}{2\pi^2} \mu_i G^2 {\rm I_3} \,,\\
  \label{e:PDET}
  a_T & = & - \frac{C_f}{4}\left(T^2-\frac{\mu_i^2}{\pi^2}\right) {\rm I_1} \,,\\
  \label{e:PDEmu}
  a_{\mu_i} & = & - \frac{1}{12}\left(\left[2N_c+N_f\right]T^2 - 
  \frac{N_cN_f}{\pi^2}\mu_i^2\right) {\rm I_2} 
  - \frac{C_f}{4}\left(T^2-\frac{\mu_i^2}{\pi^2}\right) {\rm I_3} \,,
 \end{eqnarray}
 where $C_f=(N_c^2-1)/(2N_c)$ and 
 the integral expressions explicitly read 
 \begin{eqnarray}
   {\rm I_1} & = & \frac{d_q}{2\pi^2T}\int_0^\infty dk \frac{k^2}{\omega_q} 
   \frac{\left(e^{\omega_q/T}\sin (\mu_i/T) - e^{3\omega_q/T}\sin (\mu_i/T)\right)}
   {\left(e^{2\omega_q/T}+2e^{\omega_q/T}\cos (\mu_i/T)+1\right)^2} \,,\\
   {\rm I_2} & = & \frac{d_g}{\pi^2T}\int_0^\infty dk \frac{k^2}{\omega_g}
   \frac{1}{(e^{\omega_g/T}-1)}\left(1-\frac{(\frac{4}{3}k^2+m_\infty^2)}{2\omega_g^2}
   \left[1+\frac{\omega_g}{T}\frac{e^{\omega_g/T}}{(e^{\omega_g/T}-1)}\right]\right)
   \,,\\
   \nonumber
   {\rm I_3} & = & \frac{d_q}{2\pi^2T}\int_0^\infty dk \frac{k^2}{\omega_q} 
   \bigg(\frac{2e^{\omega_q/T}\cos (\mu_i/T)+2}
   {e^{2\omega_q/T}+2e^{\omega_q/T}\cos (\mu_i/T)+1}\left[1-
   \frac{(\frac{4}{3}k^2+M_\infty^2)}{2\omega_q^2}\right] \\ \nonumber 
   & & - \frac{(\frac{4}{3}k^2+M_\infty^2)}{2\omega_qT}
   \frac{\left(2e^{3\omega_q/T}\cos (\mu_i/T)+4e^{2\omega_q/T}+
   2e^{\omega_q/T}\cos (\mu_i/T)\right)}
   {\left(e^{2\omega_q/T}+2e^{\omega_q/T}\cos (\mu_i/T)+1\right)^2} \\
   & & + \frac{\mu_i}{T}
   \frac{\left(e^{\omega_q/T}\sin (\mu_i/T) - e^{3\omega_q/T}\sin (\mu_i/T)\right)}
   {\left(e^{2\omega_q/T}+2e^{\omega_q/T}\cos (\mu_i/T)+1\right)^2}
   \bigg) \,.
 \end{eqnarray}
 The term in $b$ proportional to $a$ stems from assuming 
 temperature dependent quark masses $m_q=aT$ as employed in some 
 lattice QCD performances, e.~g.~\cite{Allton,Lombardo1,MPL,Engels,KarschEoS}. 
 
 The quark number susceptibility $\chi$ in Eq.~(\ref{e:susc}) is 
 found to be symmetric when replacing $\mu_i$ by $-\mu_i$ because 
 the same holds true for 
 $\partial G^2/\partial\mu_i$. From Eq.~(\ref{equ:PDE}) we find 
 \begin{equation}
   \label{equ:PDEApp1}
   \frac{\partial G^2}{\partial\mu_i} = \frac{b}{a_{\mu_i}} - 
   \frac{a_T}{a_{\mu_i}} \frac{\partial G^2}{\partial T} \,.
 \end{equation}
 For the individual expressions entering Eq.~(\ref{equ:PDEApp1}) 
 we find ${\rm I_1} \rightarrow -{\rm I_1}$, ${\rm I_2} \rightarrow {\rm I_2}$ 
 and ${\rm I_3} \rightarrow {\rm I_3}$ for $\mu_i\rightarrow -\mu_i$ such 
 that $b \rightarrow -b$, $a_T\rightarrow -a_T$ and $a_{\mu_i} \rightarrow 
 a_{\mu_i}$. In addition, a Taylor series expansion of 
 $G^2(T,i\mu_i)$ in powers of $\mu_i$ consists only of even powers in 
 $\mu_i$~\cite{Bluhm-PLB,MBDipl} 
 such that $\partial G^2/\partial T$ is symmetric under 
 $\mu_i\rightarrow -\mu_i$. In the limit $\mu_i\rightarrow 0$, we find 
 $\partial G^2/\partial\mu_i\rightarrow 0$ as $b\rightarrow 0$, 
 $a_T\rightarrow 0$ but $a_{\mu_i}$ and $\partial G^2/\partial T$ 
 remain non-zero. 
 
 When employing the quark dispersion relation 
 Eq.~(\ref{equ:dispq2}), the coefficients of the flow equation 
 render to 
  \begin{eqnarray}
  \label{e:PDEbn}
  \nonumber
  b & = & \left[\left(\frac{C_f}{4} + 
  m_q\sqrt{\frac{C_f}{8\left(T^2-\frac{\mu_i^2}{\pi^2}\right)G^2}}\right) 2T G^2 + 
  2m_qa + 2aM_+\right]{\rm I_1} \\ & & - 
  \frac{N_cN_f}{6\pi^2} \mu_i G^2 {\rm I_2} - 
  2\frac{\mu_i}{\pi^2} G^2 \left(\frac{C_f}{4} + 
  m_q\sqrt{\frac{C_f}{8\left(T^2-\frac{\mu_i^2}{\pi^2}\right)G^2}}
  \right){\rm I_3} \,,\\
  \label{e:PDETn}
  a_T & = & - \left(T^2-\frac{\mu_i^2}{\pi^2}\right) 
  \left(\frac{C_f}{4} + 
  m_q\sqrt{\frac{C_f}{8\left(T^2-\frac{\mu_i^2}{\pi^2}\right)G^2}}
  \right) {\rm I_1} \,,\\
  \label{e:PDEmun}
  \nonumber
  a_{\mu_i} & = & - \frac{1}{12}\left(\left[2N_c+N_f\right]T^2 - 
  \frac{N_cN_f}{\pi^2}\mu_i^2\right) {\rm I_2} \\ & & 
  - \left(T^2-\frac{\mu_i^2}{\pi^2}\right) \left(\frac{C_f}{4} + 
  m_q\sqrt{\frac{C_f}{8\left(T^2-\frac{\mu_i^2}{\pi^2}\right)G^2}}
  \right) {\rm I_3} \,.
 \end{eqnarray}

 \section*{Appendix B - Parametrizing the {\boldmath$\mu_i$} dependence}
 
\begin{figure}[t]
   \vskip 5mm
   \includegraphics[scale=0.3,angle=0.]{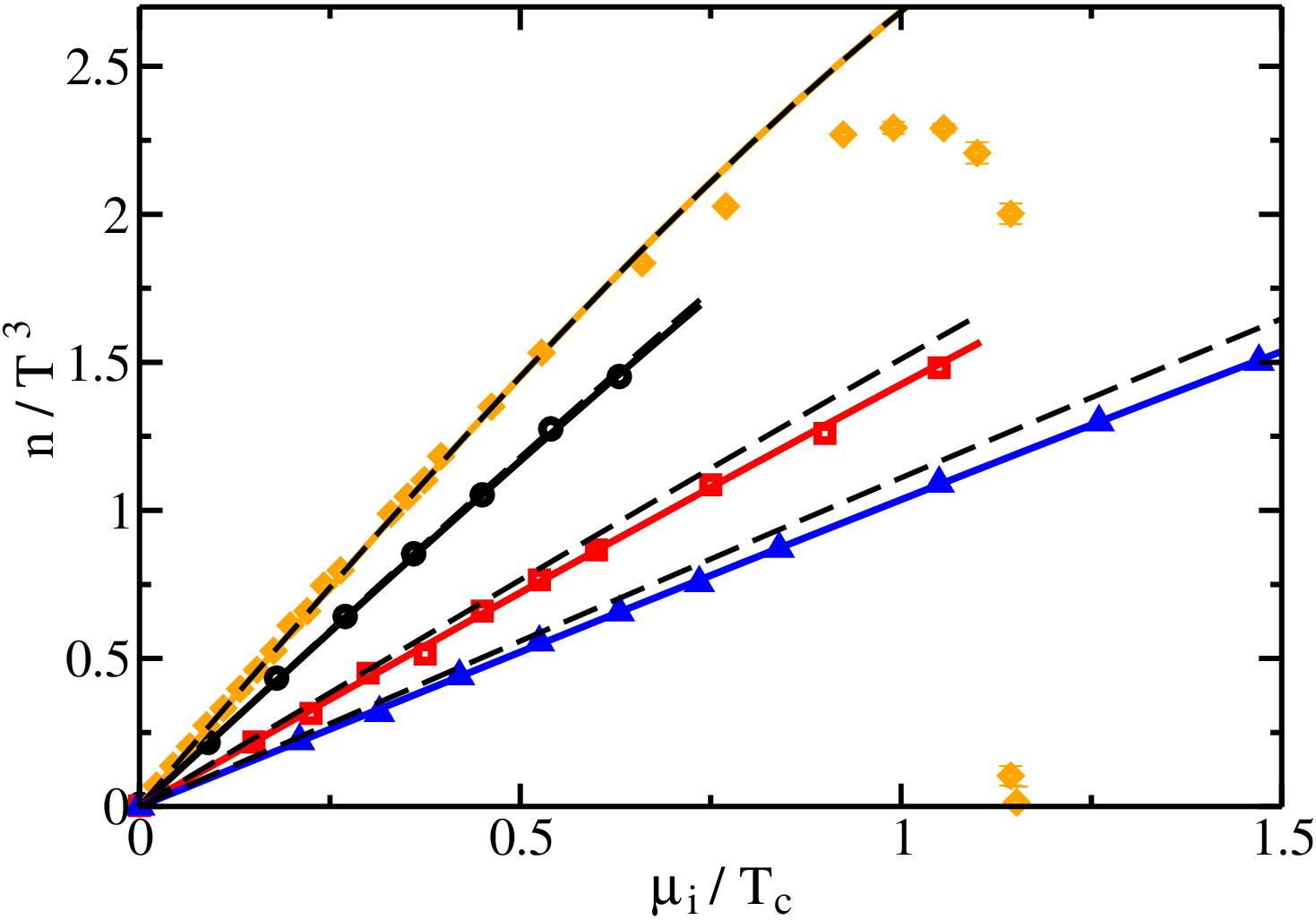}
   \hskip 3mm
   \includegraphics[scale=0.3,angle=0.]{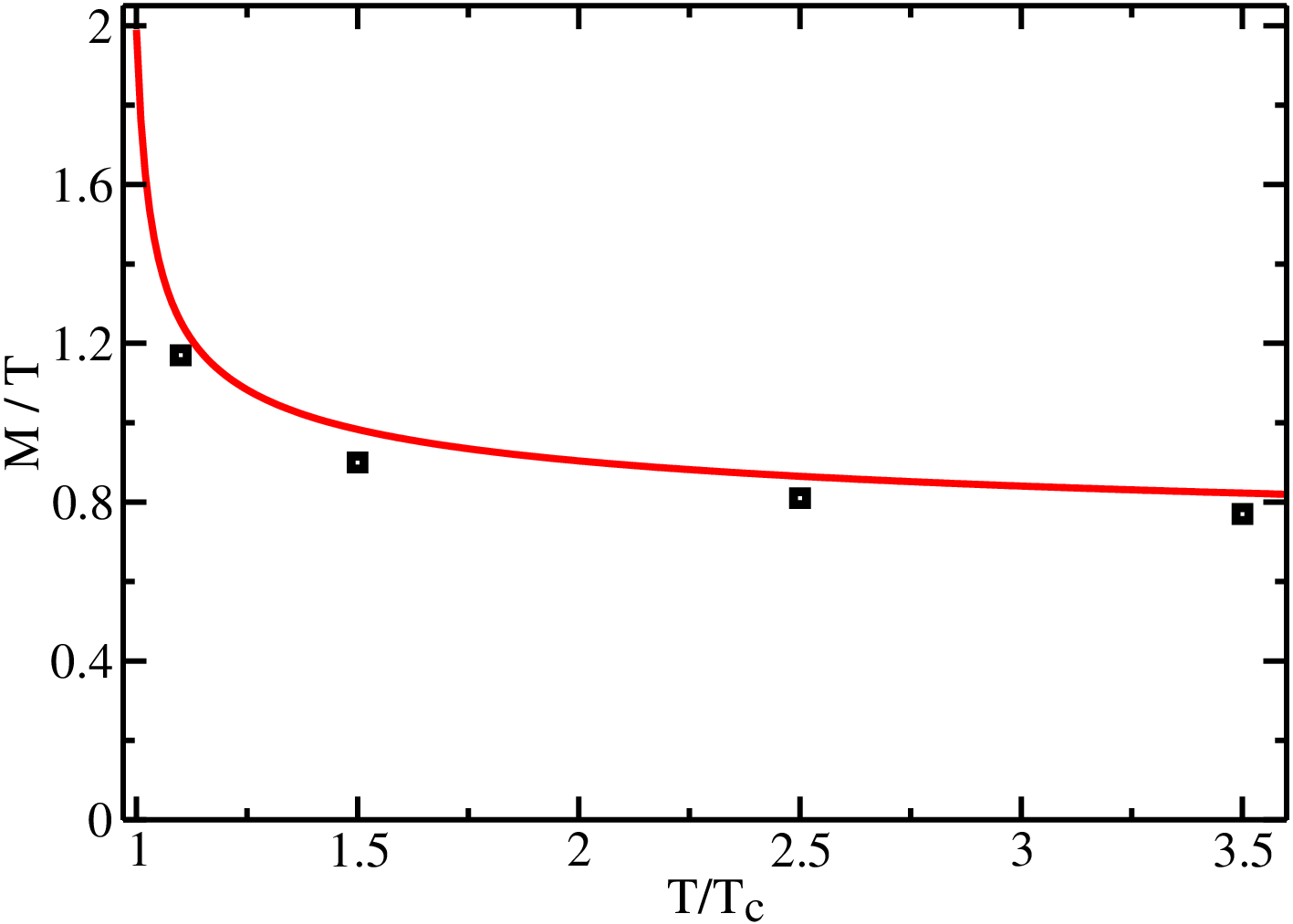}
   \caption{Left: comparison of ideal gas results for 
   $n/T^3$ as a function of $\mu_i/T_c$ 
   employing either a constant 
   mass parameter $M=0.21$ GeV (dashed curves) or readjusting 
   $M/T=1.17, 0.90, 0.81, 0.77$ for $T=1.1, 1.5, 2.5, 3.5\,T_c$ 
   (solid curves from top to bottom) with the continuum 
   extrapolated lattice QCD data (symbols) as exhibited 
   in Fig.~\ref{fig.1}. Right: comparison of found 
   $M/T$ (squares) as a function of 
   $T/T_c$ with the asymptotic quark mass $M_\infty /T$ 
   of the QPM at $\mu=0$ employing $T_s=0.96\,T_c$ and 
   $\lambda =56$ as in section~\ref{sec:3_1}. \label{fig.B1}}
 \end{figure}

\begin{figure}[t]
   \includegraphics[scale=0.35,angle=0.]{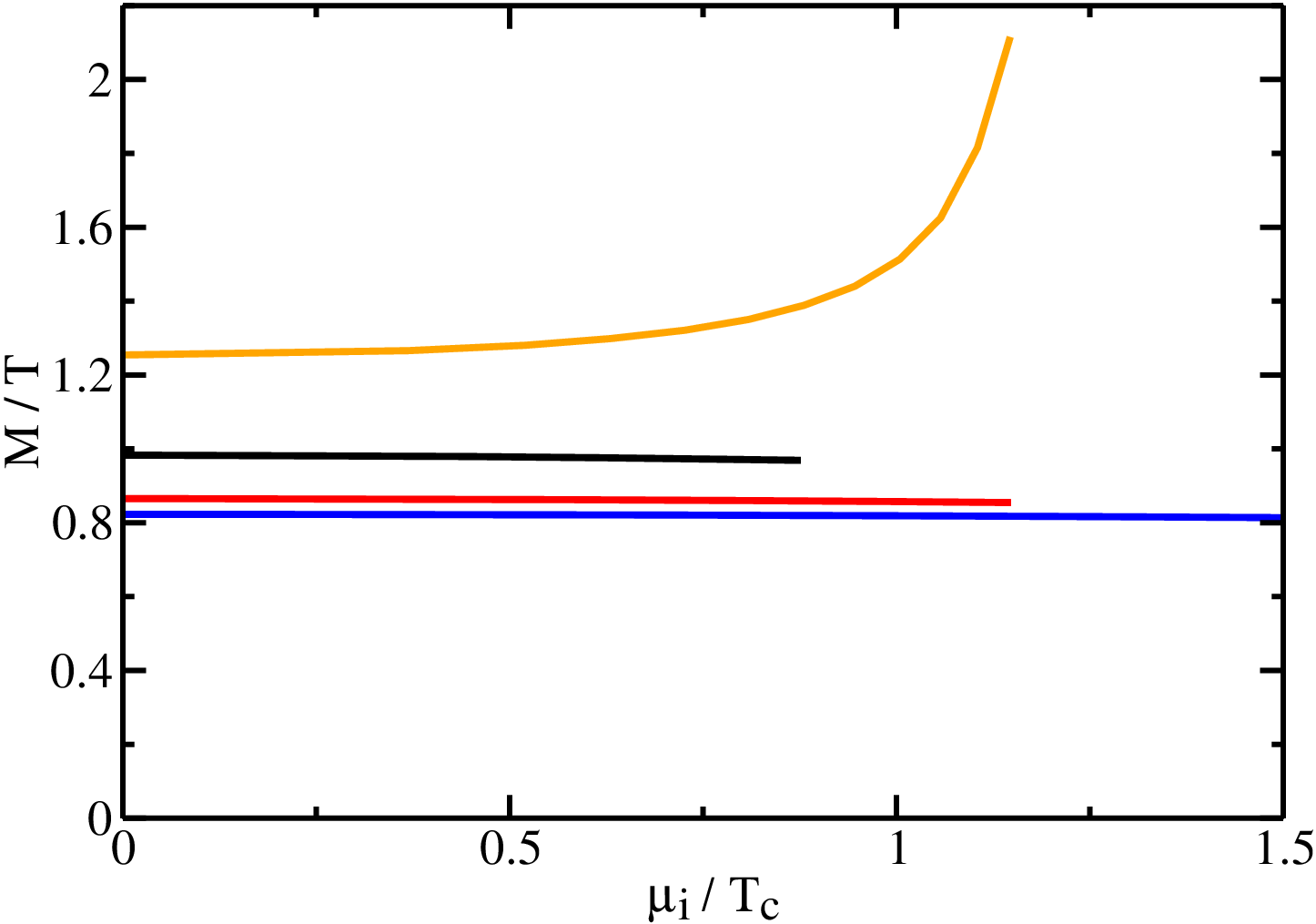}
   \caption{Asymptotic quark mass $M_\infty /T$ of the QPM 
   (solid curves) as a function of $\mu_i/T_c$ for 
   $T=1.1, 1.5, 2.5, 3.5\,T_c$ (from top to bottom) employing 
   the QPM parameters from section~\ref{sec:3_1}. \label{fig.mass2}}
 \end{figure}

 In section~\ref{sec:3}, we found a general dependence of 
 the quasiparticle dispersion relations on temperature and 
 chemical potential to be of utmost importance for 
 the successful description of lattice QCD data. 
 This shall be illustrated in some more detail by 
 considering the net quark number density in Eq.~(\ref{equ:dens1}) 
 of an ideal gas with dispersion relation $\omega_q^2=k^2+M^2$. 
 In principle, thermodynamic self-consistency demands either a dependence 
 of $M$ on both, $T$ and $\mu_i$, or neither a $T$ nor a $\mu_i$ 
 dependence of $M$. In the latter case of constant $M$, 
 we adjust $M=0.21$ GeV in order to describe the 
 continuum extrapolated lattice QCD data of $n/T^3$ 
 (cf.~Fig.~\ref{fig.1}) at $T=1.1\,T_c$ for small 
 $\mu_i/T_c$. The corresponding QPM results for 
 $T=1.1, 1.5, 2.5, 3.5\,T_c$ are then 
 exhibited in the left panel of Fig.~\ref{fig.B1} 
 (dashed curves).  
 At $T=1.1\,T_c$, we find increasing deviations from the lattice QCD 
 data for $\mu_i/T_c > 0.66$, in particular 
 in the vicinity of the Roberge-Weiss critical chemical 
 potential $\mu_c$, where the pronounced curvature cannot 
 be reproduced. This was already discussed in~\cite{MPL} 
 by considering the ratio $n(\mu_i)/n(\mu_i)_{free}$ 
 signalling clear deviations of the lattice QCD data from a free 
 (ideal) gas behavior. Furthermore, by increasing $T$ but 
 keeping $M$ fixed, the description of the lattice QCD 
 data becomes less and less accurate for smaller 
 $\mu_i/T_c$ suggesting 
 a general dependence of $M$ on $T$. Readjusting $M$ individually 
 for each temperature, ignoring for the moment being 
 thermodynamic self-consistency, the results 
 are depicted by solid curves in the left panel 
 of Fig.~\ref{fig.B1}. The found scaled mass parameters 
 $M/T$ for the temperatures considered here are exhibited 
 in the right panel of Fig.~\ref{fig.B1} (squares) and 
 compared with the scaled asymptotic quark mass 
 $M_\infty /T$ of the QPM at 
 $\mu=0$ (solid curve) employing the parametrization 
 of section~\ref{sec:3_1}. Both results agree 
 fairly well, indicating that non-zero chemical potential 
 effects are tiny for small $\mu_i/T$ but become sizeable 
 close to $\mu_c(T)$ as also visualized in Fig.~\ref{fig.mass2}. 
 In Fig.~\ref{fig.mass2}, the scaled asymptotic quark mass 
 $M_\infty /T$ of the QPM is exhibited as a function of 
 $\mu_i/T_c$ for constant $T$ using the QPM parametrization 
 of section~\ref{sec:3_1} perfectly describing $n/T^3$ 
 in Fig.~\ref{fig.1}. For increasing $T$, $M_\infty /T$ shows 
 decreasing sensitivity on $\mu_i$ while non-zero chemical 
 potential effects become important close to $\mu_c=\frac \pi 3 T$.


\newpage
 \begin{thebibliography}{100}
 
 \bibitem{Karsch} F.~Karsch, Lect. Notes Phys. {\bf 583}, 209 (2002). 
 \bibitem{Hatsuda} T.~Hatsuda, Invited Talk given at the Workshop on the 
  Physics of High Baryon Density, December 15-16, 2005, GSI, Darmstadt, Germany; 
  T.~Hatsuda, and T.~Kunihiro, Phys. Rev. Lett. {\bf 55}, 158 (1985). 
 \bibitem{POSCPOD} Proceedings of Critical Point and Onset of Deconfinement - 3rd 
  International Workshop, July 3-6, 2006, Florence, Italy, (Ed.) F.~Becattini; 
  4th International Workshop, July 9-13, 2007, Darmstadt, Germany, (Eds.) 
  P.~Senger et al.. 
 \bibitem{Gyulassy} M.~Gyulassy, and L.~McLerran, Nucl. Phys. A {\bf 750}, 30 (2005). 
 \bibitem{expdata} D.~A.~Teaney, Phys. Rev. C {\bf 68}, 034913 (2003); 
  J. Phys. G {\bf 30}, S1247 (2004); Nucl. Phys. A {\bf 785}, 44 (2007). 
 \bibitem{SQGP} E.~V. Shuryak, and I.~Zahed, Phys. Rev. D {\bf 70}, 054507 (2004);
  E.~V.~Shuryak, Nucl. Phys. A {\bf 750}, 64 (2005); 
  A.~Peshier, and W.~Cassing, Phys. Rev. Lett. {\bf 94}, 172301 (2005). 
 \bibitem{Laermann} E.~Laermann, and O.~Philipsen, Ann. Phys. Nucl. Part. Sci. 
  {\bf 53}, 163 (2003).
 \bibitem{Philipsen} O.~Philipsen, PoS {\bf LAT2005}, 016 (2006); 
  PoS {\bf JHW2005}, 012 (2006). 
 \bibitem{C_Schmidt} C.~Schmidt, PoS {\bf LAT 2006}, 021 (2006). 
 \bibitem{Peshier1} A.~Peshier, B.~K\"ampfer, O.~P.~Pavlenko, and G.~Soff, 
  Phys. Lett. B {\bf 337}, 235 (1994); 
  Phys. Rev. D {\bf 54}, 2399 (1996).
 \bibitem{Peshier3} A.~Peshier, B.~K\"ampfer, and G.~Soff, 
  Phys. Rev. C {\bf 61}, 045203 (2000); 
  Phys. Rev. D {\bf 66}, 094003 (2002). 
 \bibitem{Bluhm-EPJC} M.~Bluhm, B.~K\"ampfer, R.~Schulze, and D.~Seipt, 
  Eur. Phys. J. C {\bf 49}, 205 (2007).
 \bibitem{MBDipl} M.~Bluhm, Diploma Thesis, Technische Universit\"at Dresden, 
  August 2004. 
 \bibitem{Bluhm-PLB} M.~Bluhm, B.~K\"ampfer, and G.~Soff, 
  Phys. Lett. B {\bf 620}, 131 (2005). 
 \bibitem{RW} A.~Roberge, and N.~Weiss, Nucl. Phys. B {\bf 275}, 734 (1986). 
 \bibitem{MPL2} M.~D'Elia, and M.-P.~Lombardo, Phys. Rev. D {\bf 67}, 014505 (2003). 
 \bibitem{Seipt_Diploma} D.~Seipt, Diploma Thesis, Technische Universit\"at Dresden, 
  May 2007. 
 \bibitem{Shuryak} J.~Liao, and E.~V.~Shuryak, Phys. Rev. D {\bf 73}, 014509 (2006). 
 \bibitem{Lombardo1} M.~D'Elia, and M.-P.~Lombardo, Phys. Rev. D {\bf 70}, 074509 (2004). 
 \bibitem{MPL3} M.-P.~Lombardo, Prog. Theor. Phys. Suppl. {\bf 153}, 26 (2004); 
  M.~D'Elia, F.~Di~Renzo, and M.-P.~Lombardo, AIP Conf. Proc. {\bf 806}, 
  245 (2006). 
 \bibitem{MPL} M.~D'Elia, F.~Di~Renzo, and M.-P.~Lombardo, [{\tt arXiv:0705.3814 [hep-lat]}]. 
 \bibitem{MPL-X} M.-P.~Lombardo, Prog. Theor. Phys. Suppl. {\bf 153}, 26 (2004). 
 \bibitem{latSU3} G.~Boyd, J.~Engels, F.~Karsch, E.~Laermann, C.~Legeland, 
 M.~L\"utgemeier, and B.~Petersson, Nucl. Phys. B {\bf 469}, 419 (1996). 
 \bibitem{Karsch_overview} F.~Karsch, Lect. Notes Phys. {\bf 583}, 209 (2002). 
 \bibitem{Szabo} K.~K.~Szabo, and A.~I.~Toth, J. High Energy Phys. {\bf 06}, 008 (2003). 
 \bibitem{Gavai} R.~V.~Gavai, and S.~Gupta, Phys. Rev. D {\bf 67}, 034501 (2003); 
 (private communication, March 2006). 
 \bibitem{Lombardo3} M.-P.~Lombardo, PoS {\bf CPOD2006}, 003 (2006). 
 \bibitem{Pisarski} R.~D.~Pisarski, Nucl. Phys. A {\bf 498}, 423c (1989); 
 A.~Peshier, TFT'98 Proceedings, hep-ph/9809379.  
 \bibitem{Bluhm-PRC} M.~Bluhm, B.~K\"ampfer, R.~Schulze, D.~Seipt, and U.~Heinz, 
   Phys. Rev. C {\bf 76}, 034901 (2007). 
 \bibitem{lattice} C.~R.~Allton, S.~Ejiri, S.~J.~Hands, O.~Kaczmarek, F.~Karsch, 
   E.~Laermann, C.~Schmidt, and L.~Scorzato, Phys. Rev. D {\bf 66}, 074507 (2002). 
 \bibitem{Allton} C.~R.~Allton, S.~Ejiri, S.~J.~Hands, O.~Kaczmarek, F.~Karsch, E.~Laermann, 
  and C.~Schmidt, Phys.~Rev.~D {\bf 68}, 014507 (2003); 
  C.~R.~Allton, M.~D\"oring, S.~Ejiri, S.~J.~Hands, O.~Kaczmarek, F.~Karsch, 
  E.~Laermann, and K.~Redlich, Phys.~Rev.~D {\bf 71}, 054508 (2005). 
 \bibitem{Fodor} Z.~Fodor, S.~D.~Katz, and K.~K.~Szabo, Phys. Lett. B {\bf 568}, 73 (2003); 
   F.~Csikor, G.~I.~Egri, Z.~Fodor, S.~D.~Katz, K.~K.~Szabo, and 
   A.~I.~Toth, Nucl. Phys. Proc. Suppl. {\bf 119}, 547 (2003). 
 \bibitem{Engels} J.~Engels, R.~Joswig, F.~Karsch, E.~Laermann, M.~L\"utgemeier, and 
 B.~Petersson, Phys. Lett. B {\bf 396}, 210 (1997). 
 \bibitem{Karsch_Kitazawa} F.~Karsch, and M.~Kitazawa, [{\tt arXiv:0708.0299 [hep-lat]}]. 
 \bibitem{Blaizot} J.~P.~Blaizot, E.~Iancu, and A.~Rebhan, 
   Phys. Rev. Lett. {\bf 83}, 2906 (1999);
   Phys. Lett. B {\bf 470}, 181 (1999); Phys. Rev. D {\bf 63}, 065003 (2001);
   Phys. Lett. B {\bf 523}, 143 (2001); Phys. Rev. D {\bf 68}, 025011 (2003);
   and in {\it Quark Gluon Plasma 3}, edited by R.~C.~Hwa and X.~N.~Wang 
   (World Scientific, Singapore, 2004), p. 60.
 \bibitem{Gorenstein} M.~I.~Gorenstein, and S.~N.~Yang, Phys. Rev. D {\bf 52}, 5206 (1995).
 \bibitem{KarschEoS} F.~Karsch, E.~Laermann, and A.~Peikert, Phys. Lett. B {\bf 478}, 447 (2000). 
 \end{thebibliography}
 \end{document}